\documentclass{aa}
\usepackage{graphicx}
\usepackage{gensymb}
\usepackage{txfonts}
\usepackage{bm}
\usepackage{natbib}
\usepackage[colorlinks,
            linkcolor=red, %%修改此处为你想要的颜色
            anchorcolor=green, %%修改此处为你想要的颜色
            citecolor=blue, %%修改此处为你想要的颜色，例如修改blue为red
            ]{hyperref}
\begin{document}

\title{Dynamical essence of the eccentric von Zeipel--Lidov--Kozai effect in restricted hierarchical planetary systems}
   \author{Hanlun Lei \inst{1}\& Yan-Xiang Gong \inst{2}}
   \titlerunning{Eccentric ZLK effect}
   \authorrunning{Lei \& Gong}

   \institute{
   School of Astronomy and Space Science and Key Laboratory of Modern Astronomy and Astrophysics in Ministry of Education, Nanjing University, Nanjing 210023, China\\
   \email{leihl@nju.edu.cn}\\
   \and College of Physics and Electronic Engineering, Taishan University, Taian 271000, China
         }
   \date{Received; accepted}

% \abstract{}{}{}{}{}
% 5 {} token are mandatory

  \abstract
   {}
   {The eccentric von Zeipel--Lidov--Kozai (ZLK) effect is widely used to explain dynamical phenomena in varieties of astrophysical systems. The purpose of this work is to make clear the dynamical essence of the eccentric ZLK effect by constructing an inherent connection between such an effect and dynamics of secular resonance in restricted hierarchical planetary systems.
   }
   {Dynamical structures of apsidal resonance are analytically studied by means of perturbative treatments. The resonant model is formulated by averaging the Hamiltonian (up to octupole order) over rotating ZLK cycles, producing an additional motion integral. The phase portraits under the resonant model can be used to analyse dynamical structures, including resonant centres, dynamical separatrices and islands of libration.
   }
   {By analysing phase portraits, five branches of libration centres and eight libration zones are found in the eccentricity--inclination space. There is an excellent agreement between analytical results of libration zone and numerical distributions of resonant orbit, indicating that the resonant model for apsidal resonances is valid and applicable. Additionally, it is found that, in the test-particle limit, distributions of flipping orbits are dominated by those apsidal resonances centred at the inclination of $i=90^{\circ}$.
   }
   {The eccentric ZLK effect is dynamically equivalent to the effect of apsidal resonance in restricted hierarchical planetary systems. The dynamical response of the eccentric ZLK effect (or effect of apsidal resonance) is to significantly excite eccentricities and/or inclinations of test particles in the very long-term evolution.}
   \keywords{planets and satellites: dynamical evolution and stability --
             methods: analytical
               }

   \maketitle
%
%-------------------------------------------------------------------

\section{Introduction}
\label{Sect1}

Hierarchical three-body configurations are common in varieties of astrophysical systems, ranging from the satellite and planet scales to supermassive black hole \citep{naoz2016eccentric}. Under the test-particle approximation, such a configuration reduces to the so-called restricted hierarchical three-body problem, where the test particle moves around the central body under the gravitational perturbation from the perturber. When the perturber moves on a circular orbit, \citet{kozai1962secular} and \citet{lidov1962evolution} studied long-term dynamics of test particles and they found that a resonance occurs between the longitude of pericentre $\varpi$ and the longitude of ascending node $\Omega$ when the inclination is greater than $39.2^{\circ}$. In the long-term evolution, coupled oscillations between eccentricity and inclination are called the standard Kozai--Lidov oscillations (or Kozai--Lidov effect) \citep{lithwick2011eccentric}. About the same issue, \citet{ito2019lidov} pointed out that von Zeipel (1910) performed a similar analysis; thus they suggested to refer to such a mechanism as the von Zeipel--Lidov--Kozai (ZLK) effect.

Under the circular assumption (for the orbit of perturber), the vertical angular momentum $H=\sqrt{1-e^2}\cos{i}$ is conserved in the long-term evolution, showing that the orbits of test particle cannot flip between prograde and retrograde. It means that the standard ZLK effect cannot lead to the phenomenon of orbit flips. However, the situation is different when the circular assumption is relaxed. In this context, the Hamiltonian needs to be formulated up to the octupole order in the semimajor axis ratio. Under the octupole-level approximation, the vertical angular momentum is no longer constant \citep{naoz2016eccentric}. In particular, long-timescale modulations of Kozai--Lidov cycle can force the variation of vertical angular momentum $H$, leading to striking features including flipping between prograde and retrograde, extremely high eccentricities and chaotic behaviours \citep{katz2011long}. Such a mechanism can be naturally referred to as the eccentric ZLK effect \citep{ito2019lidov}, which have been used to interpret various dynamical phenomena in astrophysical systems \citep{libert2009kozai, shevchenko2016lidov, naoz2016eccentric}. In the test-particle limit, \citet{antognini2015timescales} investigated timescales of ZLK oscillations at quadrupole and octupole-order dynamical models.

\citet{naoz2011hot} found that planets' orbits can flip between prograde and retrograde with respect to the invariant plane of system due to the eccentric ZLK effect and they proposed a possible clue for forming hot Jupiters on retrograde orbits by combining the eccentric ZLK effect with tidal friction at the late stage. To analytically understand the eccentric ZLK effect, \citet{katz2011long} performed an average for the secular equations of motion over the period of Kozai--Lidov cycle and found a new constant of motion. In particular, they derived an analytical criterion for orbit flips. At the same time, \citet{lithwick2011eccentric} performed a numerical investigation about the eccentric ZLK effect and they numerically map out the initial conditions where flipping orbits occur for various values of $\epsilon$ ($\epsilon$ stands for the contribution of the octupole-order Hamiltonian). By using surfaces of section and Lyapunov exponents, \citet{li2014chaos} studied the chaotic and quasi-periodic evolutions caused by the eccentric ZLK effect. For the same topic, \citet{li2014eccentricity} classified flipping orbits into two types: the low-eccentricity, high-inclination (LeHi) case and high-eccentricity, low-inclination (HeLi) case. They pointed out that the first-type of flipping orbit is governed by the joint effect of the quadrupole-order and octupole-order resonances and the second type of flipping orbit is dominated by the octupole-order resonances \citep{li2014chaos, li2014eccentricity}. Recently, \citet{sidorenko2018eccentric} interpreted the eccentric ZLK effect working in the low-eccentricity, high-inclination space as a resonant phenomenon. In a recent work \citep{lei2022systematic}, a systematical study is performed for the dynamics of orbit flips caused by eccentric ZLK effect through three approaches: Poincar\'e surfaces of section, dynamical system theory (periodic orbit and invariant manifold), and perturbative treatments. Through these studies, the dynamical essence of flipping orbits is very clear: flipping orbits are a kind of quasi-periodic (or resonant) trajectory around stable, polar, periodic orbits \citep{sidorenko2018eccentric, lei2022systematic}.

However, the dynamical essence of the eccentric ZLK effect is not clear. We know that the eccentric ZLK effect is the dynamical response under the octupole-level Hamiltonian model. It means that there must be a certain correspondence between the secular dynamics and the eccentric ZLK effect. Based on this consideration, the purpose of this work is twofold. The first one is to analytically explore the dynamical structures of secular resonances (apsidal resonances) under the octupole-level approximation by means of perturbative treatments developed by \citet{henrard1986perturbation} and \citet{henrard1990semi}. This theory was adopted by \citet{sidorenko2018eccentric} to study the same topic. The second one is to construct the dynamical connection between the eccentric ZLK effect and apsidal resonances at the octupole-level approximation in restricted hierarchical planetary systems. Our results show that (a) the webs of apsidal resonance constitute basic backbones imbedded in the phase space, governing the very long-term dynamics of particles, (b) in the test-particle limit, the eccentric ZLK oscillations are attributed to the effect of apsidal resonance, and (c) only those apsidal resonances with centres at $i=90^{\circ}$ may cause orbit flips.

Through this study, the inherent connection between the eccentric ZLK effect and apsidal resonances becomes clear: from the viewpoint of dynamics, the eccentric ZLK effect is equivalent to the effect of apsidal resonance under the octupole-level approximation in restricted hierarchial planetary systems. The dynamical consequence of the eccentric ZLK effect (or effect of apsidal resonance) is to significantly excite eccentricities and/or inclinations of test particles in the long-term evolution. In particular, the behaviour of orbit flip is just one kind of dynamical response due to the eccentric ZLK effect (or effect of apsidal resonance). In this sense, the present work can be considered as an extension about the resonant interpretation for the eccentric ZLK effect \citep{sidorenko2018eccentric}.

The remaining part of this work is organised as follows. In Section \ref{Sect2}, the Hamiltonian model is briefly introduced under the test-particle and octupole-order approximation. In Section \ref{Sect3}, the fundamental frequencies and nominal location of apsidal resonance are identified under the quadrupole-order Hamiltonian flow. Resonant model for apsidal resonances is formulated in Section \ref{Sect4} by means of first-order perturbation theory. Results including the dynamical structures, libration zones and applications are presented in Section \ref{Sect5}. Finally, conclusions of this work are summarised in Section \ref{Sect6}.

\section{Hamiltonian model}
\label{Sect2}

In this work, secular resonances are investigated for an inner test particle moving around a central star under the gravitational perturbation from a distant planet\footnote{The planet acts the role of perturber.}. Such a dynamical model is called restricted hierarchical planetary system, which is widely adopted as the basic dynamical model to study secular dynamics in varieties of astrophysical systems \citep{lithwick2011eccentric, li2014chaos, li2014eccentricity, sidorenko2018eccentric, luo2016double, lei2018modified, lei2019semi, katz2011long, antognini2015timescales, lei2021structures, lei2021, lei2022systematic}. The mass of the central star is denoted by $m_0$ and the mass of the perturber is denoted by $m_p$. In the test-particle limit, the orbit of the perturber around the central star is unchanged, while the test particle moves around the central star on a perturbed Keplerian orbit. Under such a hierarchical configuration, the invariant plane of system is coincident with the orbit of perturber.

For convenience, let us introduce a right-handed inertial reference frame, with the origin at the central star, $x$--$y$ plane at the invariable plane (i.e., the perturber's orbit), $x$-axis along the eccentricity vector of the perturber's orbit and $z$-axis parallel to the vector of the total angular momentum. Under such a coordinate system, the orbits of test particle (perturber) are described by the semimajor axis $a (a_p)$, the eccentricity $e (e_p)$, inclination $i (i_p)$, longitude of ascending node $\Omega (\Omega_p)$, argument of pericentre $\omega (\omega_p)$ and mean anomaly $M(M_p)$. For both prograde and retrograde configurations, the longitude of pericentre and mean anomaly can be defined in a general manner \citep{shevchenko2016lidov}
\begin{equation*}
\varpi = \Omega + {\rm sign} (\cos{i}) \omega, \quad \lambda = M + \varpi
\end{equation*}
for the test particle and
\begin{equation*}
\varpi_p = \Omega_p + {\rm sign} (\cos{i_p}) \omega_p,\quad \lambda_p = M_p + \varpi_p
\end{equation*}
for the perturber. Here ${\rm sign} (x)$ is a sign function of $x$ and it is equal to $1.0$ when $x$ is greater than zero and it is equal to $-1.0$ when $x$ is smaller than zero. Under the chosen reference frame, it holds $i_p = 0$ and $\varpi_p = 0$\footnote{This is due to the setting that the $x$-axis is along the eccentricity vector of the perturber's orbit.}. Without otherwise stated, in the entire work we adopt the variables with subscript $p$ for the perturber and the ones without subscripts for test particle.

In the long-term evolution, the short-period terms arising in the Hamiltonian can be filtered out by means of double-averaging techniques over the orbital periods of the test particle and the perturber \citep{ford2000secular, naoz2013secular, naoz2016eccentric, luo2016double, shevchenko2016lidov, lei2018modified, lei2019semi}. Such a process of phase averaging is called secular approximation \citep{naoz2016eccentric}. Due to the hierarchial configuration, the semimajor axis ratio $\alpha = \frac{a}{a_p}$ is a small parameter, leading to the fact that the Hamiltonian can be truncated at a certain order in semimajor axis ratio\footnote{The Hamiltonian truncated at the second order corresponds to the quadrupole-level approximation, and the one truncated at the third order corresponds to the octupole-level approximation.}.

The (normalised) double-averaged Hamiltonian, up to the octupole order in the semimajor axis ratio, can be written as \citep{lithwick2011eccentric, naoz2016eccentric}
\begin{equation}\label{Eq1}
\begin{aligned}
{\cal H} =  - \left( {{F_{\rm quad}} + \epsilon {F_{\rm oct}}} \right)
\end{aligned}
\end{equation}
where the coefficient $\epsilon$, measuring the significance of the octupole-order contribution, is a small parameter, given by
\begin{equation*}
\epsilon  = \frac{a}{{{a_p}}}\frac{{{e_p}}}{{1 - e_p^2}}
\end{equation*}
showing that the semimajor axis ratio $\alpha = \frac{a}{{{a_p}}}$ or the eccentricity of the perturber $e_p$ is larger, the contribution of the octupole-order term is greater. The quadrupole-order term is given by
\begin{equation*}
{F_{\rm quad}} =  - \frac{1}{2}{e^2} + {\cos ^2}i + \frac{3}{2}{e^2}{\cos ^2}i + \frac{5}{2}{e^2}\left( {1 - {{\cos }^2}i} \right)\cos \left( {2\omega } \right)
\end{equation*}
and the octupole-order term is given by
\begin{equation*}
\begin{aligned}
{F_{\rm oct}} = &\frac{5}{{16}}\left( {e + \frac{3}{4}{e^3}} \right)\\
&\times \left[ {\left( {1 - 11\cos i - 5{{\cos }^2}i + 15{{\cos }^3}i} \right)\cos \left( {\omega  - \Omega } \right)} \right.\\
&\left. { + \left( {1 + 11\cos i - 5{{\cos }^2}i - 15{{\cos }^3}i} \right)\cos \left( {\omega  + \Omega } \right)} \right]\\
&- \frac{{175}}{{64}}{e^3}\left[ {\left( {1 - \cos i - {{\cos }^2}i + {{\cos }^3}i} \right)\cos \left( {3\omega  - \Omega } \right)} \right.\\
&\left. { + \left( {1 + \cos i - {{\cos }^2}i - {{\cos }^3}i} \right)\cos \left( {3\omega  + \Omega } \right)} \right]
\end{aligned}
\end{equation*}
The double-averaged Hamiltonian up to an arbitrary order in $\alpha$ can be found in \citet{laskar2010explicit} and \citet{lei2021}.

In order to formulate the Hamiltonian model, a set of (normalised) Delaunay variables are introduced as follows \citep{lithwick2011eccentric}:
\begin{equation*}
\begin{aligned}
g &= \omega ,\quad G = \sqrt {1 - {e^2}}, \\
h &= \Omega ,\quad H = G\cos i.
\end{aligned}
\end{equation*}
In terms of Delaunay's variables, the Hamiltonian can be further expressed as follows:
\begin{equation}\label{Eq2}
\begin{aligned}
{\cal H} \left(g,h,G,H\right) &= {\cal H}_2 \left(g,G,H\right) + {\cal H}_3 \left(g,h,G,H\right)\\
&= - F_{\rm quad} \left(g,G,H\right) - \epsilon F_{\rm oct} \left(g,h,G,H\right),
\end{aligned}
\end{equation}
which determines a dynamical model with two degrees of freedom. Hamiltonian canonical relations lead to the equations of motion as follows \citep{morbidelli2002modern}:
\begin{equation}\label{Eq3}
\begin{aligned}
\frac{{{\rm d}g}}{{{\rm d} t}} &= \frac{{\partial {\cal H}}}{{\partial G}},\quad \frac{{{\rm d}G}}{{{\rm d} t}} =  - \frac{{\partial {\cal H}}}{{\partial g}},\\
\frac{{{\rm d}h}}{{{\rm d} t}} &= \frac{{\partial {\cal H}}}{{\partial H}},\quad \frac{{{\rm d}H}}{{{\rm d} t}} =  - \frac{{\partial {\cal H}}}{{\partial h}}.
\end{aligned}
\end{equation}
The Hamiltonian given by Eq. (\ref{Eq2}) holds the following symmetries \citep{sidorenko2018eccentric}:
\begin{equation*}
{\cal H}\left( g,h,G,H\right) = {\cal H}\left( 2\pi - g,h,G,-H\right) = {\cal H}\left( g,2\pi - h,G,-H\right),
\end{equation*}
which implies that the solution curves under the Hamiltonian flow are symmetric with respect to $H=0$ (i.e., $i=90^{\circ}$). It is noted that there is a unique parameter $(\epsilon)$ that characterizes the dynamical model. The effectiveness of the standard double-averaging process requires that $\epsilon$ should be a small parameter ($\epsilon \ll 1$) and the mass of the perturber should be much smaller than that of the central star ($m_p \ll m_0$) \citep{naoz2016eccentric, luo2016double, lei2018modified, lei2019semi}. For all the following simulations, the dynamical model with system parameter $\epsilon = 0.03$ is adopted as an example\footnote{Without doubt, the method adopted in this work is applicable to dynamical models specified by other values of $\epsilon$.}.

\section{Nominal location of apsidal resonance}
\label{Sect3}

In this section, let us identify the fundamental frequencies under the quadrupole-order dynamical model and then it becomes possible for us to determine the nominal location of secular resonance.

The quadrupole-order Hamiltonian ${\cal H}_2$ is very simple and can be written as
\begin{equation}\label{Eq4}
\begin{aligned}
{\cal H}_2 =   &\frac{1}{2}\left( {1 - {G^2}} \right) - \frac{{{H^2}}}{{{G^2}}} - \frac{3}{2}\left( {\frac{1}{{{G^2}}} - 1} \right){H^2}\\
&- \frac{5}{2}\left( {1 - {G^2} + {H^2} - \frac{{{H^2}}}{{{G^2}}}} \right)\cos \left( {2g} \right)
\end{aligned}
\end{equation}
where the angular coordinate $h$ is absent from ${\cal H}_2$, indicating that the $z$-component of the angular momentum $H$ is conserved under the quadrupole-order model. The conserved quantity $H$ can be specified by the critical inclination $i_c$ (when the eccentricity is assumed as zero) of the manner \citep{kozai1962secular},
\begin{equation*}
 H = \sqrt{1-e^2}\cos{i} = \cos{i_c}.
\end{equation*}
In the following discussions, we often use $i_c$ to stand for $H$. The dynamical model determined by ${\cal H}_2$ is of one degree of freedom. With given $H$ (or $i_c$), the solution curves under the Hamiltonian flow of ${\cal H}_2$ are usually called the ZKL cycles, including librating ZKL cycles and rotating ZKL cycles. Rotating and librating ZKL cycles are divided by means of a dynamical separatrix in the phase space.

\begin{figure}
\centering
\includegraphics[width=0.4\textwidth]{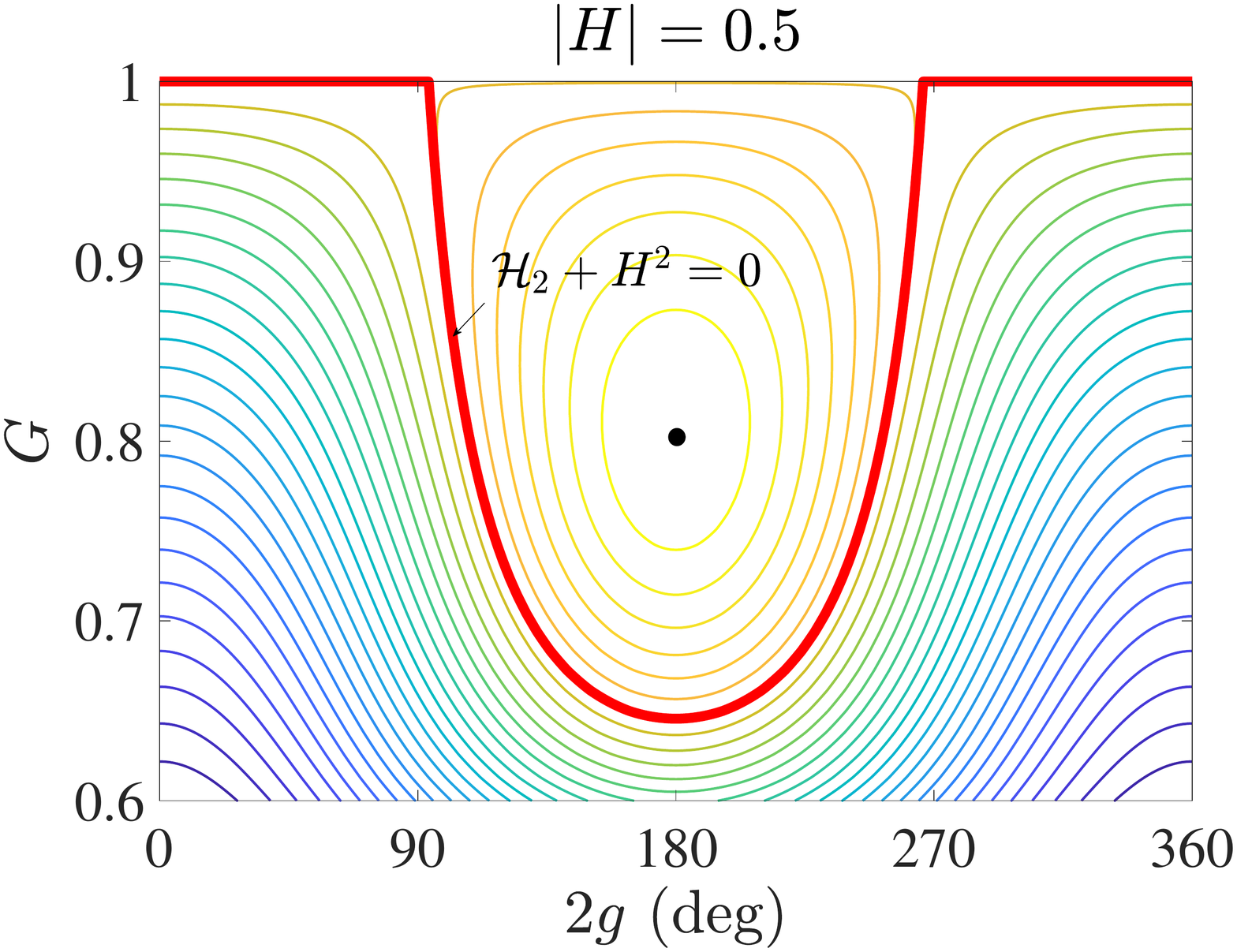}
\caption{Phase portrait (level curves of Hamiltonian in the phase space) with the motion integral at $\left|H\right| = 0.5$ (corresponding to the critical inclination at $i_c = 60^{\circ}$ or $i_c = 120^{\circ}$) under the quadrupole-order Hamiltonian model. Kozai centre is marked by black dot and dynamical separatrix is shown in red line. The regions filled with circulating and librating ZKL cycles are divided by the dynamical separatrix expressed by ${\cal H}_2 + H^2 = 0$. For the current example, the phase space with ${\cal H}_2 > -0.25$ is of ZKL libration and the space with ${\cal H}_2 < -0.25$ is of ZKL circulation.}
\label{Fig0}
\end{figure}

The global structures in the phase space can be explored by analysing phase portraits, which correspond to level curves of Hamiltonian with given motion integral. As an example, the case of $\left|H\right| = 0.5$ (corresponding to the critical inclination at $i_c = 60^{\circ}$ or $i_c = 120^{\circ}$) is considered and the associated phase portrait is presented in Fig. \ref{Fig0}. In this case, the Kozai--Lidov resonance may occur. At the ZKL centre, the eccentricity and inclination should satisfy \citep{kozai1962secular}
\begin{equation*}
\cos^2{i} = \frac{3}{5} \left(1-e^2\right).
\end{equation*}
In Fig. \ref{Fig0}, the black dot stands for the position of ZKL centre, at which the Hamiltonian takes its maximum. The dynamical separatrix, as shown in red line, divides the rotating ZKL cycles from librating ZLK cycles. It is known that the separatrix corresponds to the level curve of Hamiltonian passing through $G=1$ (corresponding to $e=0$). Substituting $G=1$ into the quadrupole-order Hamiltonian ${\cal H}_2$ given by Eq. (\ref{Eq4}), it is not difficult to get the expression of separatrix as \citep{lei2021structures}
\begin{equation*}
{\cal H}_2 +H^2 = 0.
\end{equation*}
For the current example (the motion integral is taken as $\left|H\right|=0.5$), the Hamiltonian of separatrix is equal to ${\cal H}_2 = -0.25$. From the phase portrait, we observe that, in the phase space, the region with ${\cal H}_2 > -0.25$ is of ZLK libration and the region with ${\cal H}_2 < -0.25$ is of ZLK circulation.

From Fig. \ref{Fig0}, we can further observe that the ZLK cycles starting from $2g=0$ are of circulation. In the following, we will focus on those regions filled with rotating ZLK cycles\footnote{It is noted that, inside the regions filled with ZLK librating cycles, the dynamics is dominated by the quadrupole-order resonance (i.e., ZLK resonance). Only in those regions filled with ZLK rotating cycles, the octupole-order resonance plays an important role in governing the very long-term behaviours of particles \citep{li2014chaos}.} where the octupole-order resonances may appear and dominate the long-term dynamics. Without loss of generality, we fix the initial condition at $2g=0$ for rotating ZLK cycles \citep{katz2011long, li2014chaos, sidorenko2018eccentric, lithwick2011eccentric} and then determine the fundamental frequencies in the parameter space spanned by orbit elements $(e,i)$ or conserved quantities $({\cal H}_2, H)$.

\begin{figure*}
\centering
\includegraphics[width=0.4\textwidth]{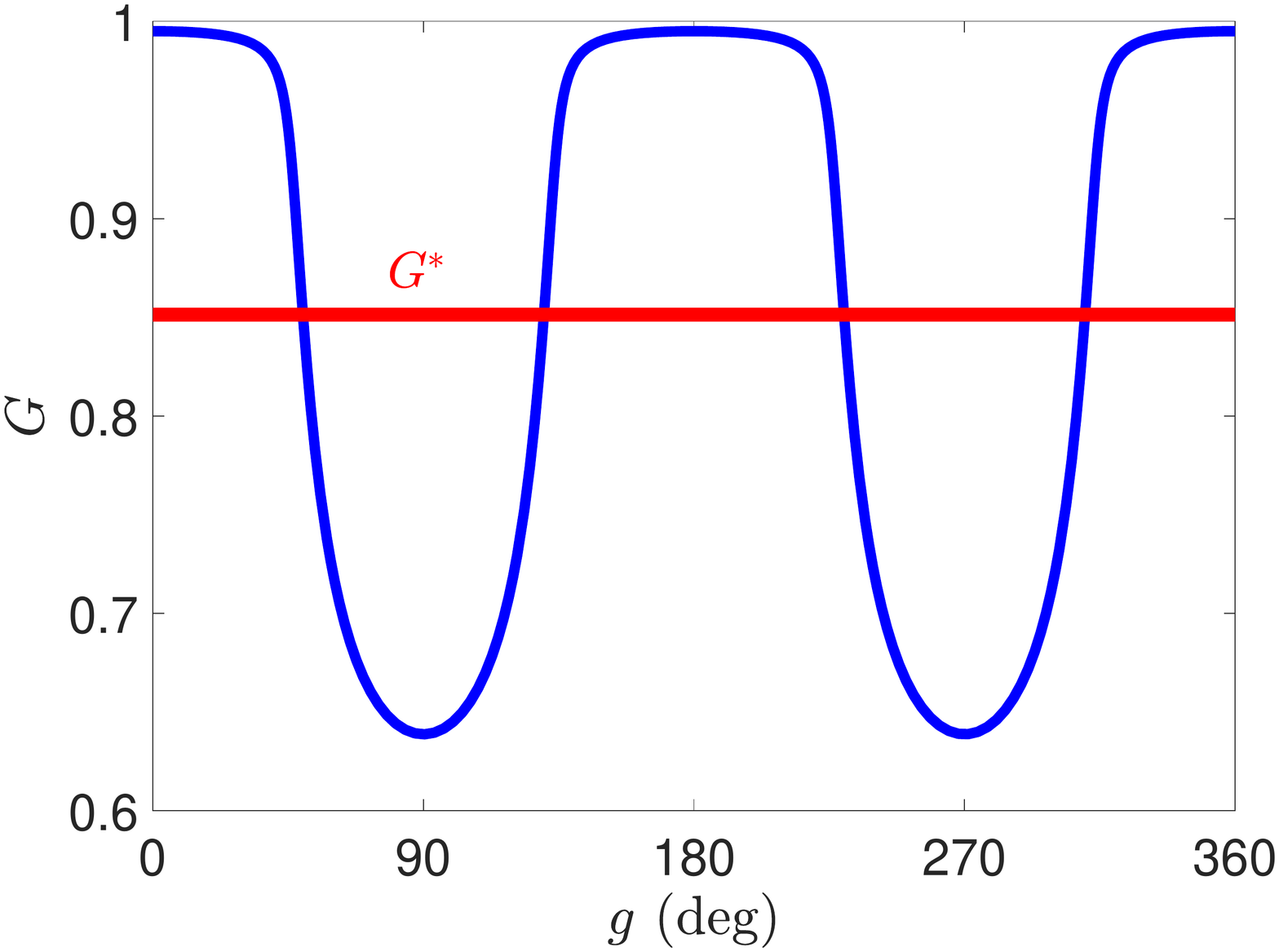}
\includegraphics[width=0.4\textwidth]{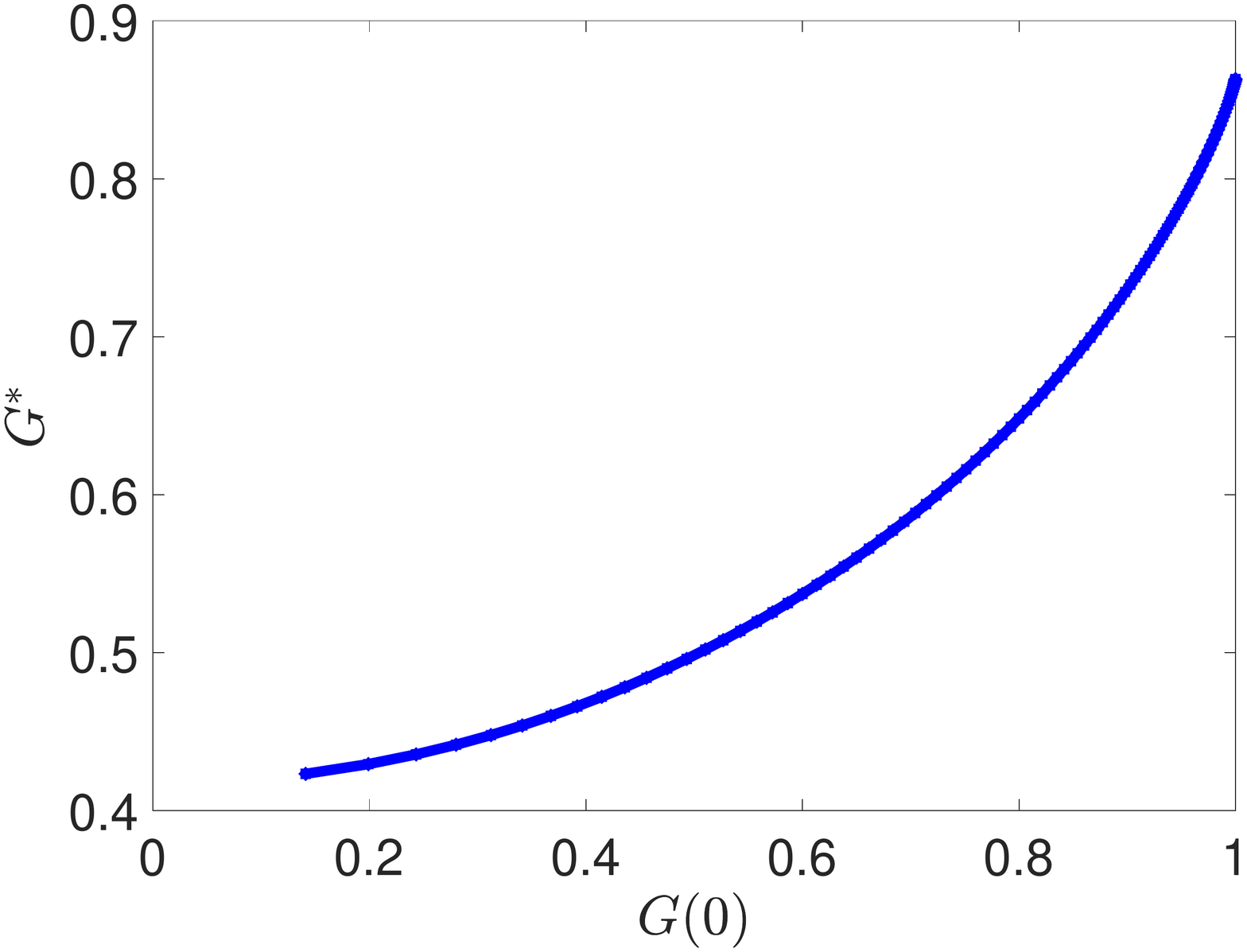}
\caption{An example of rotating ZLK cycle shown in the phase space (\emph{left panel}) and the Arnold action $G^*$ as a function of $G(0)$ (\emph{right panel}), under the quadrupole-order Hamiltonian flow. Please refer to the text for the definition of the Arnold action $G^*$. For the example shown in the left panel, the initial condition is taken as $g_0 = 0$, $h_0 = 2\pi$, $G_0 = 0.9950$ and $H_0 = 0.4975$.}
\label{Fig1}
\end{figure*}

In order to study the dynamics of secular resonance by means of perturbative treatments \citep{henrard1986perturbation, henrard1990semi}, let us introduce the following action--angle variables under the quadrupole-order Hamiltonian flow,
\begin{equation}\label{Eq5}
\begin{aligned}
{g^*} &= g - {\rho _g}\left( {{g^*},{G^*},{H^*}} \right),\quad {G^*} = \frac{1}{{2\pi }}\int\limits_0^{2\pi } {G{\rm d}g},\\
{h^*} &= h - {\rho _h}\left( {{g^*},{G^*},{H^*}} \right),\quad {H^*} = H,
\end{aligned}
\end{equation}
which is canonical with the generating function,
\begin{equation*}
S_1\left( {g,h,{G^*},{H^*}} \right) = h{H^*} + \int {G\left( {{{\cal H}_2}\left( {{G^*},{H^*}} \right),g,{H^*}} \right){\rm d}g}.
\end{equation*}
In Eq. (\ref{Eq5}), $\rho_g$ and $\rho_h$ are periodic functions with the same period of the rotating ZLK cycle. The variable ${G^*}$ is called Arnold action, which stands for the phase-space area bounded by the ZLK cycle (divided by $2\pi$).

Under the quadrupole-order dynamical model, let us denote the period of $g$ as $T_g$ and the period of $h$ as $T_h$. Thus, the linear functions of $g^*$ and $h^*$ can be expressed as
\begin{equation*}
g^* = \frac{2\pi}{T_g} t, \quad h^* = h_0^* + \frac{2\pi}{T_h} t.
\end{equation*}
At the initial instant, it holds $g^* = 0$ and $h^* = h_0^*$ for rotating ZLK cycles.

In Fig. \ref{Fig1}, a rotating ZLK cycle in the phase space $(g,G)$ is shown in the left panel and the relation between the Arnold action $G^*$ and $G(0)$ is plotted in the right panel. In the left panel, the location of $G^*$ for the particular example is marked in red. Evidently, $G^*$ is a constant under the quadrupole-order Hamiltonian flow.

In practice, we produce the ZLK cycle as well as the action $G^*$ by integrating the following differential equations over one period of ZLK cycle (i.e., $T_g$) under the quadrupole-order Hamiltonian flow,
\begin{equation*}
\dot g = \frac{{\partial {{\cal H}_2}}}{{\partial G}},\quad \dot G =  - \frac{{\partial {{\cal H}_2}}}{{\partial g}},\quad \dot W = G\frac{{\partial {{\cal H}_2}}}{{\partial G}}.
\end{equation*}
At the initial instant, it holds $g_0 = 0$ and $W_0 = 0$. Here $W(t)$ stands for the oriented area enclosed by the solution curve $G(g)$ under the quadrupole-level Hamiltonian flow. In particular, when the integration time is equal to one period of the rotating ZLK cycle, it holds
\begin{equation*}
W\left( {{T_g}} \right) = \int\limits_0^{{T_g}} {G\frac{{\partial {{\cal H}_2}}}{{\partial G}}{\rm d}t}  = \int\limits_0^{2\pi } {G{\rm d}g}  = 2\pi {G^*}.
\end{equation*}
Thus, the Arnold action $G^*$ is equal to $\frac{1}{2\pi} W\left( {{T_g}} \right)$. It is mentioned that the ZLK cycles $G(g)$ and the Arnold action $G^*$ can be alternatively produced by means of elliptic integrals, as presented by \citet{sidorenko2018eccentric}.

According to the generating function, we can obtain an alternative expression for the new set of angles $(g^*, h^*)$ as follows:
\begin{equation*}
{g^*} = \frac{{\partial S_1}}{{\partial {G^*}}},\quad {h^*} = \frac{{\partial S_1}}{{\partial {H^*}}}.
\end{equation*}
Thus, we can get the expressions for computing periodic functions $\rho_g = g - g^*$ and $\rho_h = h - h^*$ as
\begin{equation*}
\begin{aligned}
{\rho _g}\left(g^*,G^*,H^*\right) = &\int {\frac{{\partial {{\cal H}_2}}}{{\partial G}}{\rm d}t}\\
&- \frac{\partial }{{\partial {G^*}}}\int {G\left( {{{\cal H}_2}\left( {{G^*},{H^*}} \right),g,{H^*}} \right){\rm d}g},\\
{\rho _h}\left(g^*,G^*,H^*\right) =  &- \frac{\partial }{{\partial {H^*}}}\int {G\left( {{{\cal H}_2}\left( {{G^*},{H^*}} \right),g,{H^*}} \right){\rm d}g},
\end{aligned}
\end{equation*}
which indicate that ${\rho _g} = g - g^*$ and ${\rho _h} = h - h^*$ are equal to zero when $g^* = 0$ or $g^* = 2\pi$ \citep{henrard1990semi}. It means that the old and new set of angles are coincident at the initial instant and at one period of ZLK cycle.

\begin{figure*}
\centering
\includegraphics[width=0.4\textwidth]{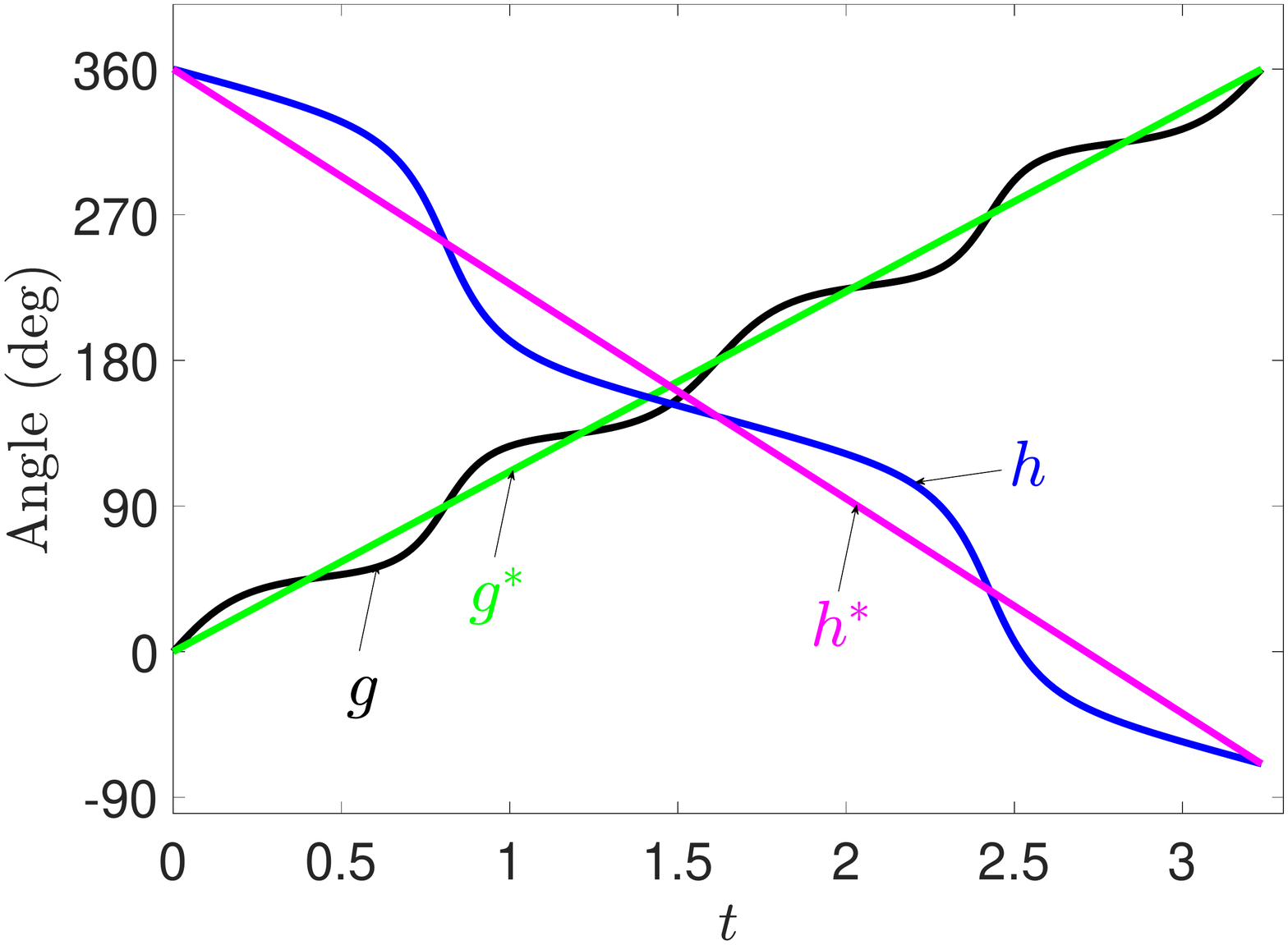}
\includegraphics[width=0.4\textwidth]{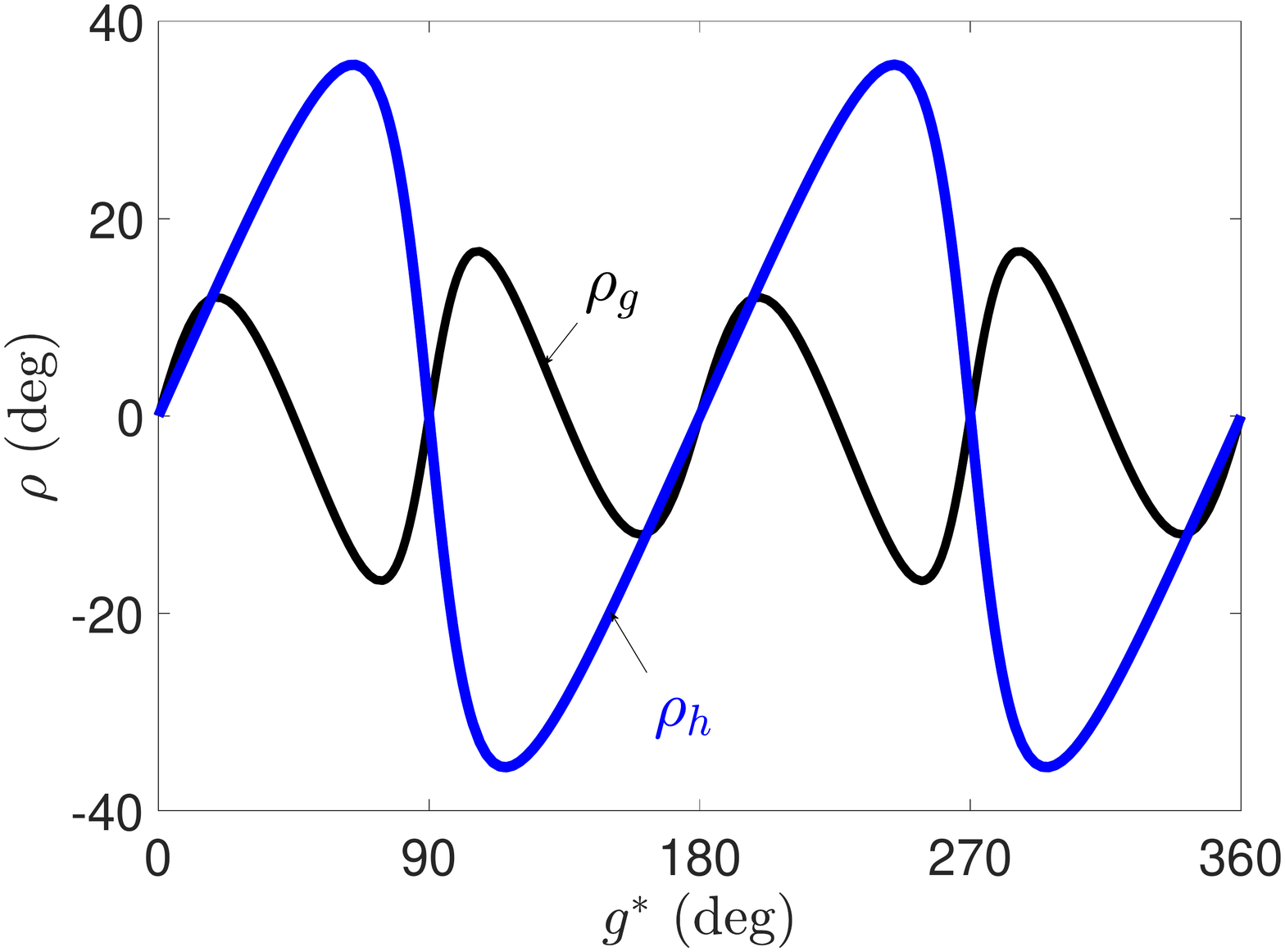}
\caption{Old and new sets of angular coordinates $(g,h)$ and $(g^*,h^*)$ as functions of time (\emph{left panel}) as well as the differences between them $\rho_g = g - g^*$ and $\rho_h = h - h^*$ as functions of $g^*$ (\emph{right panel}). These two plots correspond to the example shown in the left panel of Fig. \ref{Fig1}.}
\label{Fig2}
\end{figure*}

For the example shown in the left panel of Fig. \ref{Fig1}, the time histories of $(g,h)$ and $(g^*,h^*)$ are shown in the left panel of Fig. \ref{Fig2} and the differences between the old and new set of angles $\rho_g$ and $\rho_h$ as functions of $g^*$ are reported in the right panel of Fig. \ref{Fig2}. It is observed that (a) the new angles $g^*$ and $h^*$ are linear functions of time, and (b) $\rho_g$ and $\rho_h$ are periodic functions of $g^*$ and they are equal to zero when $g^*$ is at $0$, $\pi/2$, $\pi$, $3\pi/2$ and $2\pi$.

Under the canonical transformation given by Eq. (\ref{Eq5}), the quadrupole-level Hamiltonian ${\cal H}_2$ becomes \citep{henrard1990semi}
\begin{equation}\label{Eq6}
{\cal H}_2 \left(g, G, H\right) = {\cal H}_2 \left(G^*, H^*\right),
\end{equation}
which shows that $g^*$ and $h^*$ are absent from the Hamiltonian ${\cal H}_2$, indicating that $G^*$ and $H^*$ are conserved quantities along the ZLK cycle. The fundamental frequencies under the quadrupole-order dynamical model are identified by
\begin{equation}\label{Eq7}
{{\dot g}^*} = \frac{{\partial {{\cal H}_2}\left( {{G^*},{H^*}} \right)}}{{\partial {G^*}}},\quad
{{\dot h}^*} = \frac{{\partial {{\cal H}_2}\left( {{G^*},{H^*}} \right)}}{{\partial {H^*}}}
\end{equation}
which determines the periods of $g$ and $h$ as
\begin{equation*}
{T_g} = \frac{{2\pi }}{{{{\dot g}^*}}},\quad {T_h} = \frac{{2\pi }}{{{{\dot h}^*}}}.
\end{equation*}

\begin{figure*}
\centering
\includegraphics[width=0.4\textwidth]{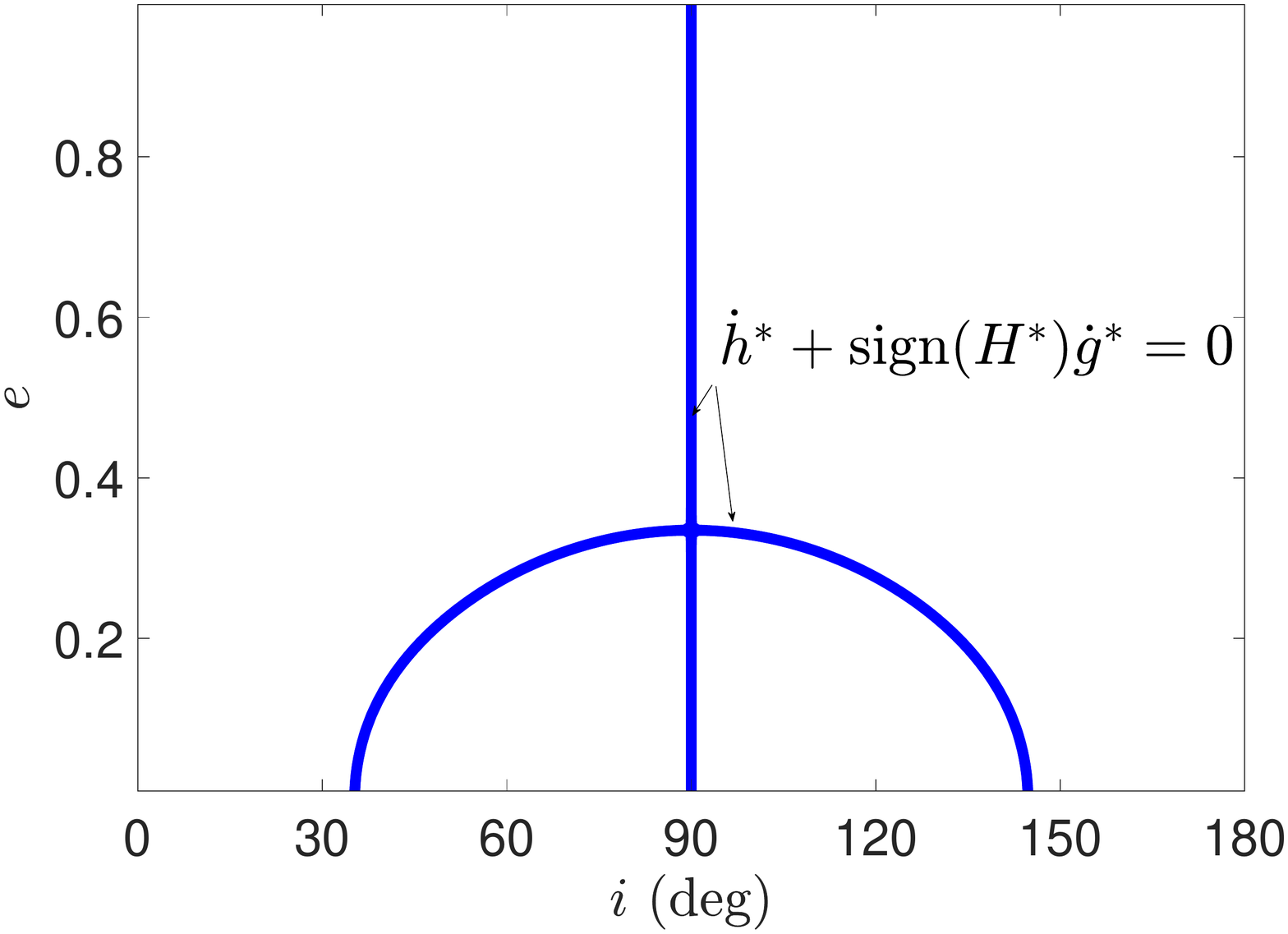}
\includegraphics[width=0.4\textwidth]{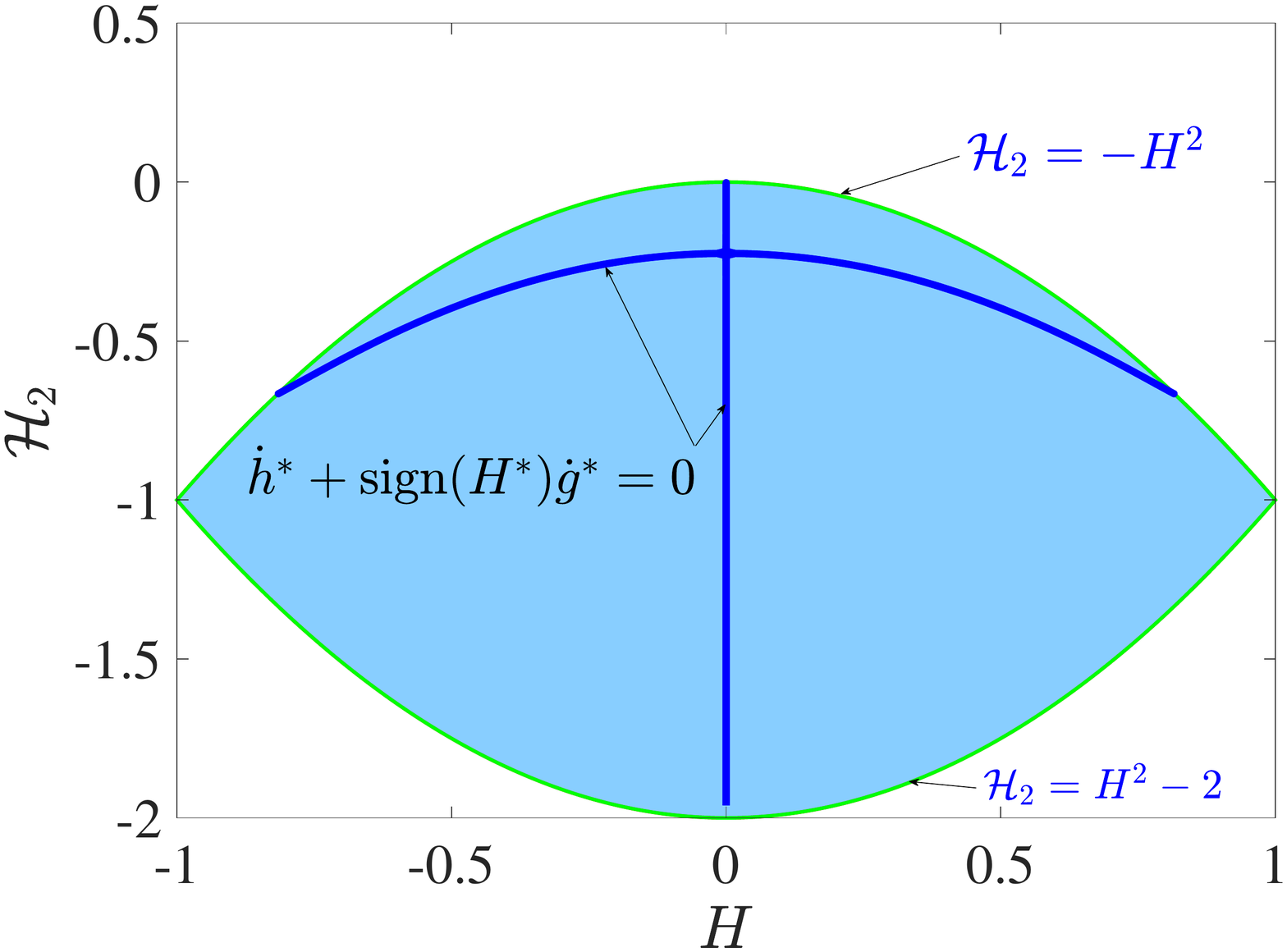}
\caption{Nominal location of apsidal resonance with critical argument of $\sigma_1 = h^* + {\rm sign}(H^*)g^*$, shown in the $(i,e)$ space (\emph{left panel}) and in the $(H, {\cal H}_2)$ space (\emph{right panel}). The shaded region in the right panel is of ZLK circulation.}
\label{Fig3}
\end{figure*}

Based on the fundamental frequencies, it is possible for us to determine the nominal location of secular resonance by the following resonant condition,
\begin{equation}\label{Eq8}
k_1 {\dot g}^* + k_2 {\dot h}^* = 0,
\end{equation}
where $k_1, k_2 \in \mathbb{Z}$.

As for the octupole-order resonance, the critical argument is given by
\begin{equation*}
\sigma = h^* + {\rm sign}(H^*) g^*.
\end{equation*}
In Fig. \ref{Fig3}, the solution of ${\dot \sigma} = {\dot h}^* + {\rm sign}(H^*) {\dot g}^* = 0$\footnote{It corresponds to $k_1 = 1$ and $k_2 = {\rm sign}(H^*)$, means that $k_2$ is related to the inclination.} is distributed in the $(i,e)$ space in the left panel and in the $(H,{\cal H}_2)$ space in the right panel. In the left panel, the eccentricity and inclination are evaluated when the angle $g$ is equal to zero\footnote{This is because we fix $2g=0$ as initial conditions of rotating ZLK cycles.}. In the right panel, the shaded region is of ZLK circulation, the upper boundary of circulation region is given by ${\cal H}_2 + H^2 = 0$ and the bottom boundary is given by ${\cal H}_2 - H^2 + 2 = 0$ \citep{lei2021structures}. In both panels, the distribution of nominal location of resonant centre is symmetric with respect to $i=90^{\circ}$ (or $H=0$).

From the left panel of Fig. \ref{Fig3}, we can observe that there are three branches of resonant centres in the considered parameter space: one branch corresponds to polar orbits at arbitrary eccentricities and the other two branches occupy the low-eccentricity space. The latter two branches are called asymmetric families of resonant centre. From the right panel of \ref{Fig3}, we can see that these two asymmetric branches are close to the upper boundary of circulation region represented by ${\cal H}_2 + H^2 = 0$ (this boundary corresponds to the ZLK separatrix, as discussed in Fig. \ref{Fig0}).

Next, let us discuss the essence of the resonance with critical argument of $\sigma = h^* + {\rm sign}(H^*) g^*$. According to the `general' definition of longitude of pericentre $\varpi^*$ for prograde and retrograde orbit configurations \citep{shevchenko2016lidov},
\begin{equation*}
\varpi^* = \Omega^* + {\rm sign}(\cos i^*) \omega^*,
\end{equation*}
we can see that the critical argument $\sigma$ is equal to the longitude of pericentre $\varpi^*$. Because of the choice of the reference frame, it is known that the longitude of pericentre of the perturber's orbit is fixed at zero, i.e., $\varpi_p = 0$ and ${\dot \varpi}_p = 0$. As a result, we can further write the critical argument as
\begin{equation}\label{Eq9}
\sigma = h^* + {\rm sign}(H^*) g^* = \varpi^* =  \varpi^* - \varpi_p
\end{equation}
which means that, in essence, the resonances arising in Fig. \ref{Fig3} are the so-called apsidal resonances with critical argument of $\sigma = \varpi^* - \varpi_p$. For simplicity, we denote the critical argument of apsidal resonance as $\sigma = h^* + {\rm sign}(H^*) g^*$ in the test-particle limit\footnote{In the test-particle limit, the longitude of the perturber with $\varpi_p = 0$ is taken into account.}. Based on the set of Delaunay variables $(g,h,G,H)$, \citet{sidorenko2018eccentric} defined the critical argument as $\sigma = h + {\rm sign}(H) g$, which is also adopted by \citet{lei2022systematic} in his study. Additionally, \citet{katz2011long} introduced the longitude $\Omega_e = \Omega + \arctan\left(\tan{\omega}\cos{i}\right)$ to describe the very long-term behaviours caused by eccentric ZLK effect. Discussions about the relation between $\Omega_e$ and $\sigma = h + {\rm sign}(H) g$ are made in \citet{lei2022systematic}.

In the coming section, we will study the dynamics of apsidal resonance from the viewpoint of perturbative treatments developed by \citet{henrard1986perturbation} and \citet{henrard1990semi}. The core concept is to consider the octupole-order term in the Hamiltonian as the perturbation to the quadrupole-order dynamics.

\section{Resonant Hamiltonian of apsidal resonance}
\label{Sect4}

In the previous section, we know that the apsidal resonances with critical argument of $\sigma = h^* + {\rm sign}(H^*) g^*$ can happen in the considered parameter space. The purpose of this section is to formulate the resonant Hamiltonian by means of first-order perturbation theory \citep{henrard1990semi}. This theory was also adopted by \citet{sidorenko2018eccentric}.

Under the new set of canonical variables $(g^*,h^*,G^*,H^*)$, the Hamiltonian up to the octupole order in $\alpha$ can be expressed as
\begin{equation}\label{Eq10}
{\cal H}\left( {{g^*},{h^*},{G^*},{H^*}} \right) =  {\cal H}_2 \left( {{G^*},{H^*}} \right) + {\cal H}_3 \left( {{g^*},{h^*},{G^*},{H^*}} \right),
\end{equation}
where the quadrupole-order Hamiltonian ${\cal H}_2$ (independent on the angular coordinates) is considered as the unperturbed part and the octupole-order Hamiltonian ${\cal H}_3$ plays the role of perturbation to the quadrupole-order dynamics\footnote{The magnitude of perturbation is measured by the small parameter $\epsilon$.}. In such a perturbed Hamiltonian model, the unperturbed part ${\cal H}_2$ is also called the kernel function, which yields the fundamental (or proper) frequencies.

In order to study the dynamics of apsidal resonance with critical argument of $\sigma = h^* + {\rm sign}(H^*) g^*$, the following transformation is introduced,
\begin{equation}\label{Eq11}
\begin{aligned}
{\sigma _1} &= {h^*} + {\rm sign}\left( {{H^*}} \right){g^*},\quad {\Sigma _1} = {H^*}\\
{\sigma _2} &= {g^*},\quad {\Sigma _2} = {G^*} - \left| {{H^*}} \right|
\end{aligned}
\end{equation}
which is canonical with the generating function,
\begin{equation*}
S_2\left( {{g^*},{h^*},{\Sigma _1},{\Sigma _2}} \right) = {h^*}{\Sigma _1} + {g^*}\left( {\left| {{\Sigma _1}} \right| + {\Sigma _2}} \right).
\end{equation*}
It is noted that a similar transformation to Eq. (\ref{Eq11}) is introduced by \citet{sidorenko2018eccentric} and \citet{lei2022systematic} but based on the set of canonical variables $(g,h,G,H)$.

Under the new set of canonical variables $(\sigma_1,\sigma_2,\Sigma_1,\Sigma_2)$, the Hamiltonian can be further written as
\begin{equation}\label{Eq12}
{\cal H}\left(\sigma_1,\sigma_2,\Sigma_1,\Sigma_2\right) =  {\cal H}_2 \left( \Sigma_1, \Sigma_2 \right) + {\cal H}_3 \left(\sigma_1,\sigma_2,\Sigma_1,\Sigma_2\right).
\end{equation}
It should be mentioned that it is difficult to obtain the explicit expression of Eq. (\ref{Eq12}). In practice, we compute the Hamiltonian ${\cal H}$ numerically once the set of variables $\left(\sigma_1,\sigma_2,\Sigma_1,\Sigma_2\right)$ is given.

At the resonant centre shown in Fig. \ref{Fig3}, it holds
\begin{equation*}
{\dot \sigma}_1 = \frac{\partial {\cal H}_2}{\partial \Sigma_1} = 0.
\end{equation*}
As the octupole-order Hamiltonian ${\cal H}_3$ is small compared to the quadrupole-order Hamiltonian ${\cal H}_2$, we can obtain
\begin{equation*}
{\dot \sigma}_1 = \frac{\partial {\cal H}}{\partial \Sigma_1} = \frac{\partial {\cal H}_2}{\partial \Sigma_1} + \frac{\partial {\cal H}_3}{\partial \Sigma_1} = \frac{\partial {\cal H}_3}{\partial \Sigma_1} \sim {\cal O}(\epsilon)
\end{equation*}
which shows that, under the perturbation of the octupole-order interaction, apsidal resonances may also take place.

When the test particle is inside an apsidal resonance, the resonant angle $\sigma_1$ becomes a long-period variable and the angle $\sigma_2$ is a short-period variable. This is a separable Hamiltonian system \citep{henrard1990semi}. Thus, the terms involving $\sigma_2$ in the Hamiltonian yield short-period influences, thus they can be filtered out by means of averaging technique\footnote{Averaging treatment corresponds to the lowest-order perturbation method \citep{naoz2016eccentric}.}. To this end, we can formulate the resonant Hamiltonian by performing a further average for the Hamiltonian over one period of $\sigma_2$,
\begin{equation}\label{Eq13}
{{\cal H}^ * }\left( {{\sigma _1},{\Sigma _1},{\Sigma _2}} \right) = \frac{1}{{2\pi }}\int\limits_0^{2\pi } {{\cal H}\left( {{\sigma _1},{\sigma _2},{\Sigma _1},{\Sigma _2}} \right){\rm d}{\sigma _2}}.
\end{equation}
Because of the definition $\sigma_2 = g^*$, we can understand that such an average is performed over one period of a rotating ZLK cycle under the quadrupole-order Hamiltonian flow \citep{sidorenko2018eccentric, katz2011long}. Under the dynamical model determined by Eq. (\ref{Eq13}), the angle $\sigma_2$ becomes a cyclic coordinate, so that its action $\Sigma_2$ becomes a constant of motion (or motion integral). The resulting resonant model determined by Eq. (\ref{Eq13}) is of one degree of freedom, depending on the motion integral $\Sigma_2$.

When the eccentricity is assumed as zero, the motion integral $\Sigma_2$ can be specified by a critical inclination $i_*$,
\begin{equation}\label{Eq14}
{\Sigma _2} = {G^*} - \left| {{H^*}} \right|= 1 - \left|\cos {i_*}\right|
\end{equation}
The critical inclination $i_*$ corresponds to the minimum inclination in the prograde space and corresponds to the maximum inclination in the retrograde space. It is not difficult to get that $i_*$ and $\pi - i_*$ stand for the same motion integral.

\section{Results}
\label{Sect5}

In the previous section, the resonant Hamiltonian for apsidal resonances is formulated as ${\cal H}^*\left(\sigma_1,\Sigma_1,\Sigma_2\right)$ where $\Sigma_2$ is the motion integral of the resonant model. The global dynamics of apsidal resonance in the phase space can be revealed by phase portraits. In this section, we produce phase portraits and then analyse the phase portraits to estimate resonant width \citep{lei2021, lei2021dynamical, lei2022systematic}. At last, we construct a connection between the libration zones of apsidal resonance and numerical distributions of resonant orbit (or flipping orbit).

\subsection{Dynamical structures of apsidal resonance}
\label{Sect5_1}

\begin{figure*}
\centering
\includegraphics[width=0.4\textwidth]{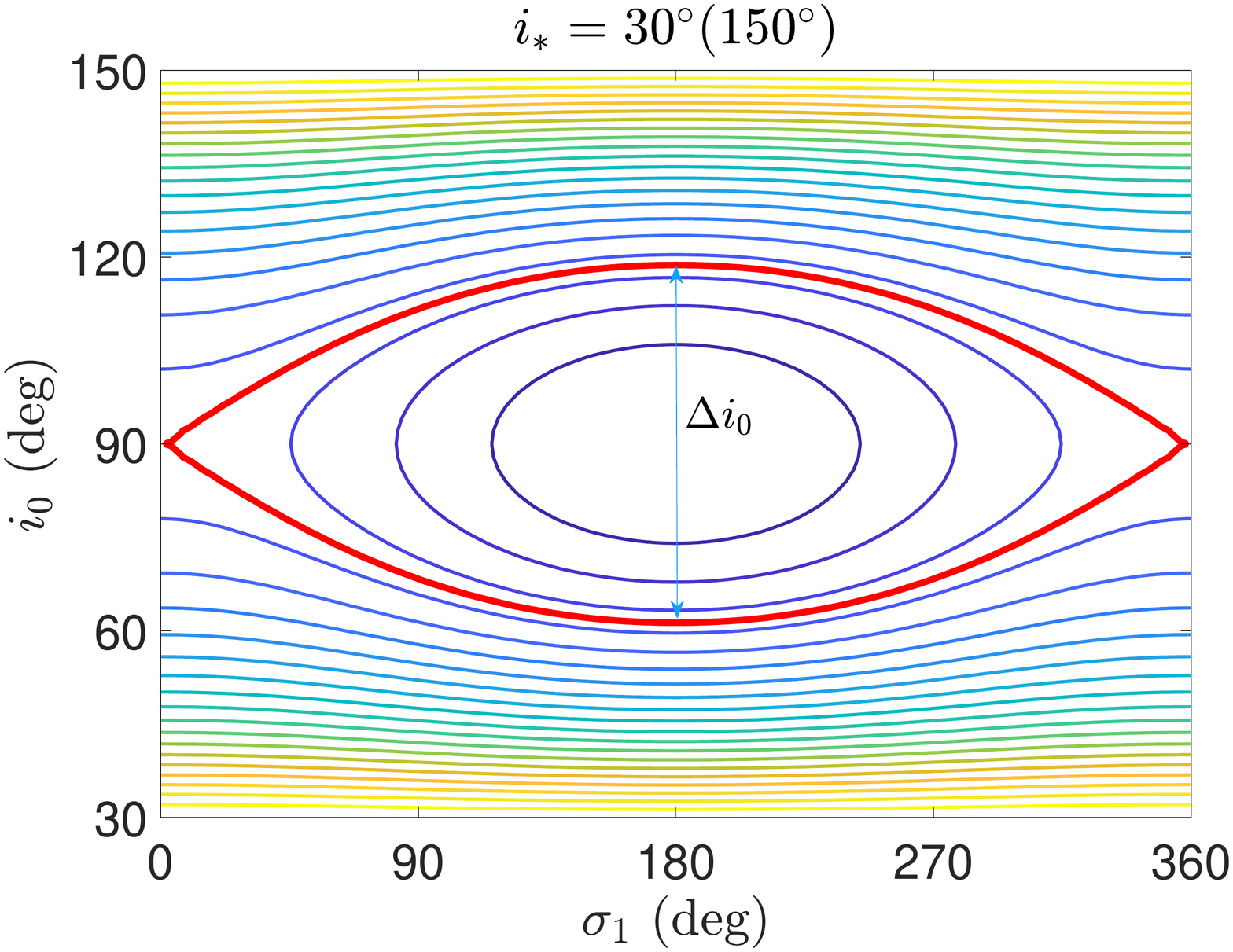}
\includegraphics[width=0.4\textwidth]{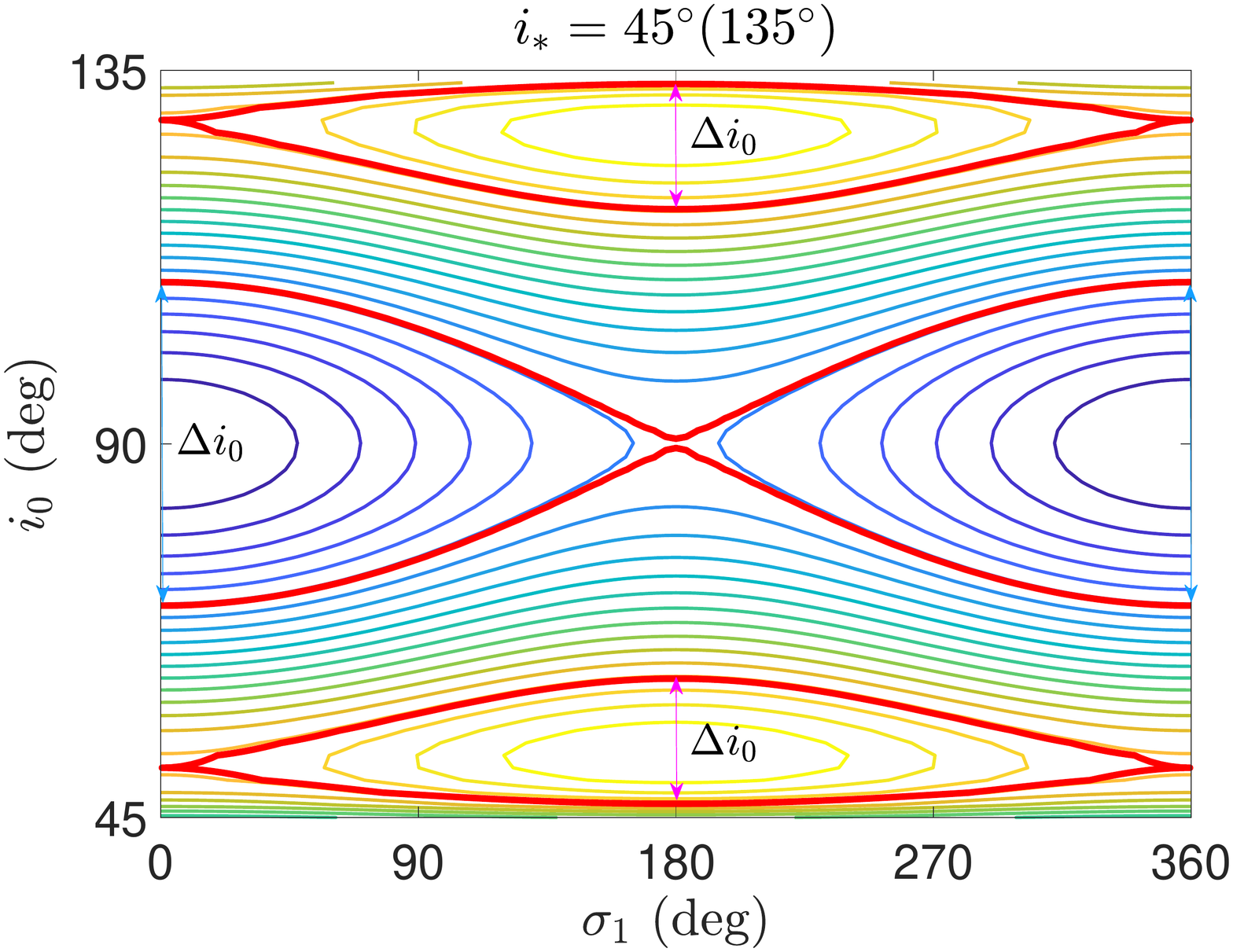}\\
\includegraphics[width=0.4\textwidth]{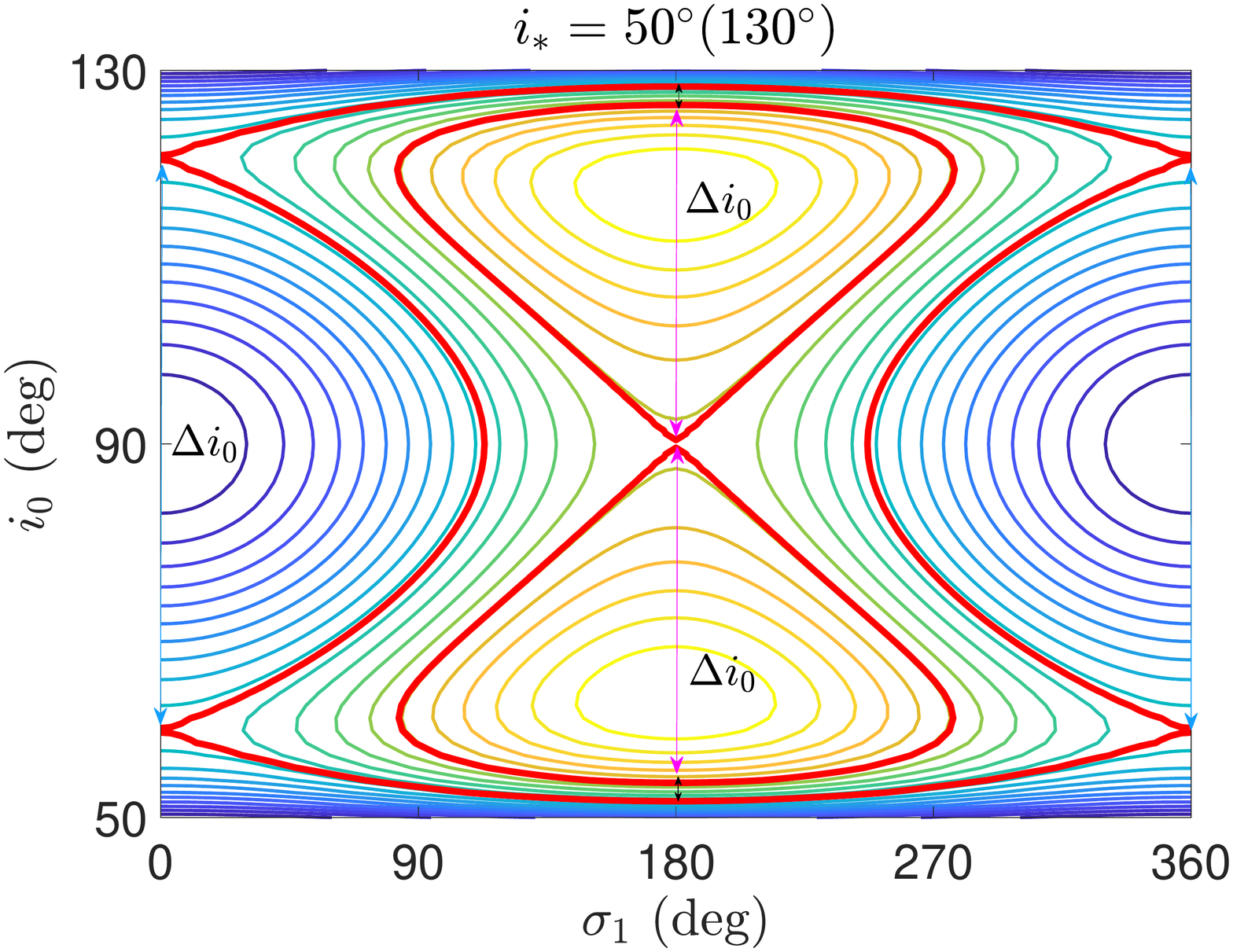}
\includegraphics[width=0.4\textwidth]{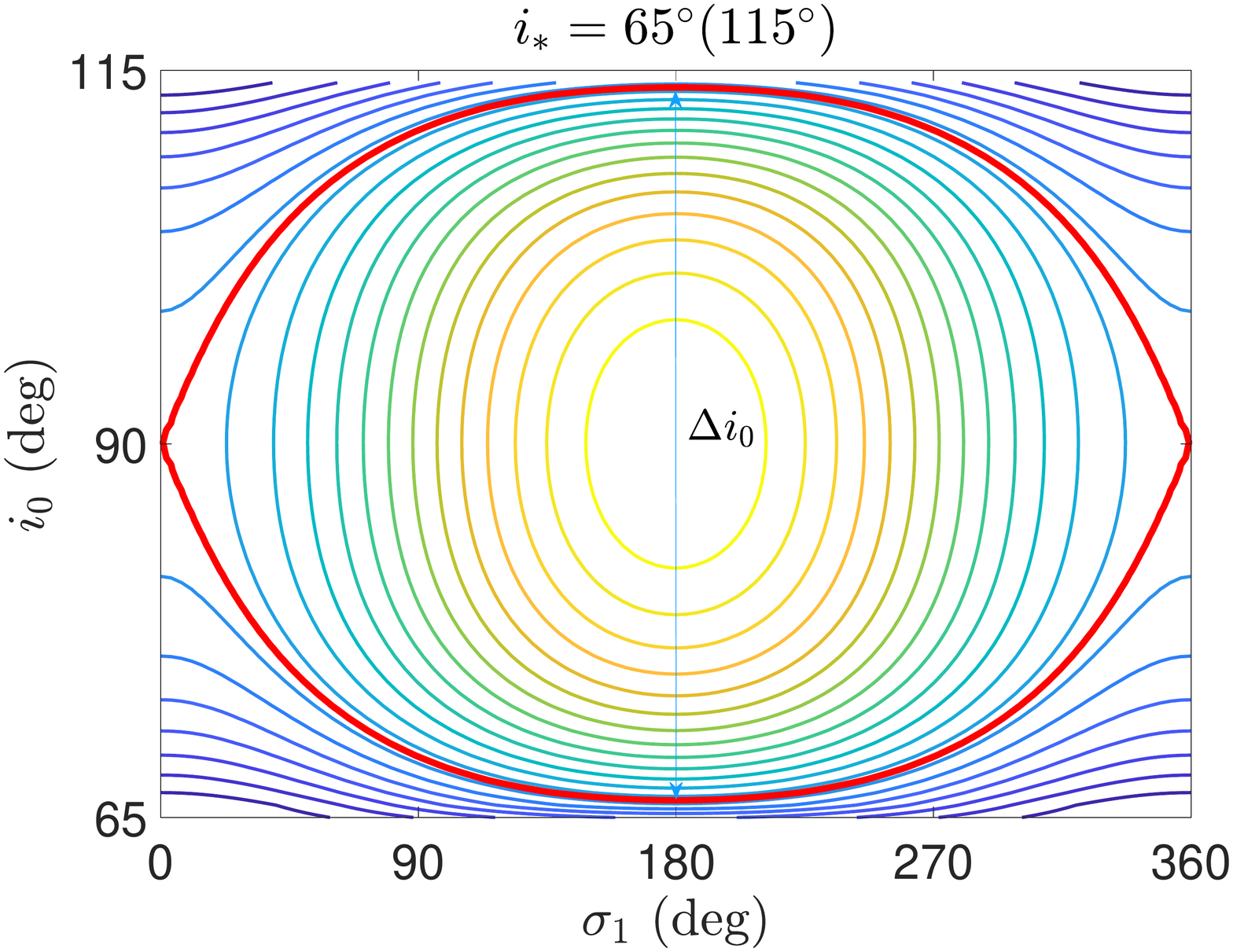}
\caption{Phase portraits (level curves of resonant Hamiltonian) of apsidal resonance with critical argument of $\sigma_1 = h^* + {\rm sign}(H^*)g^*$ at different levels of motion integral specified by $i_*$. Here $i_*$ represents the magnitude of motion integral $\Sigma_2$ (please refer to the text for details). It is noted that $i_0$ shown in the $y$-axis corresponds to the inclination when the angle $g$ is equal to zero. Dynamical separatrices are marked in red lines. The resonant width, denoted by $\Delta i_0$, measures the maximum size of island of libration.}
\label{Fig4}
\end{figure*}

In Fig. \ref{Fig4}, the (pseudo-) phase portraits are shown in the $(\sigma_1, i_0)$ space for the motion integral specified by the critical inclination at $i_* = 30^{\circ} (150^{\circ})$, $45^{\circ} (135^{\circ})$, $50^{\circ} (130^{\circ})$ and $65^{\circ} (115^{\circ})$. It should be noted that $i_0$ given in the $y$-axis corresponds to the inclination when the angle $g$ is equal to zero\footnote{This is because we assume $2g=0$ as the initial conditions for rotating ZLK cycles. This assumption is often used in this work.}. Dynamical separatrices are marked in red lines and the resonant width, measuring the maximum size of the island of libration, is denoted by $\Delta i_0$. Evidently, all the dynamical structures are symmetric with respect to the lines of $i_0 = 90^{\circ}$, due to the symmetry of the Hamiltonian function.

\begin{figure*}
\centering
\includegraphics[width=0.4\textwidth]{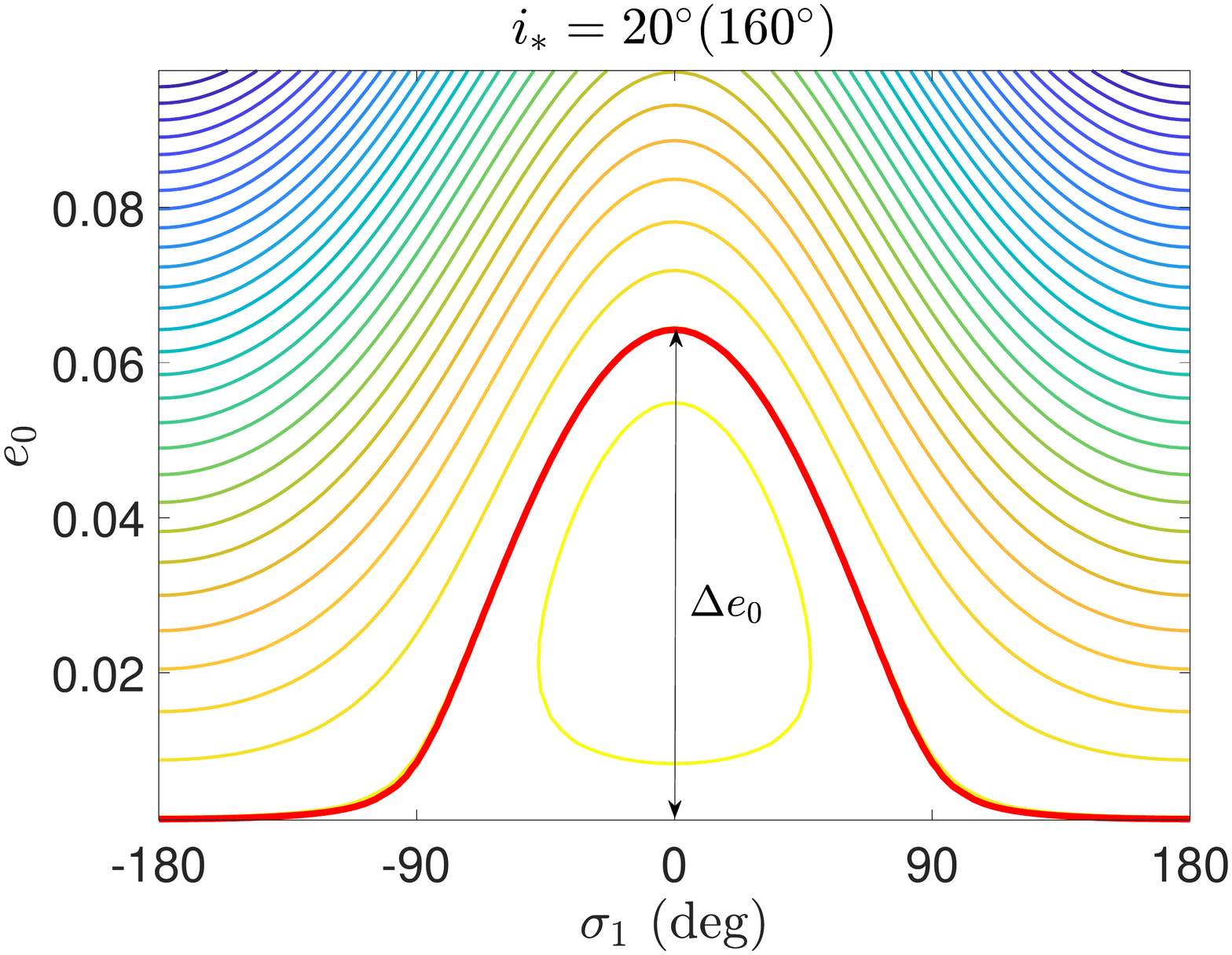}
\includegraphics[width=0.4\textwidth]{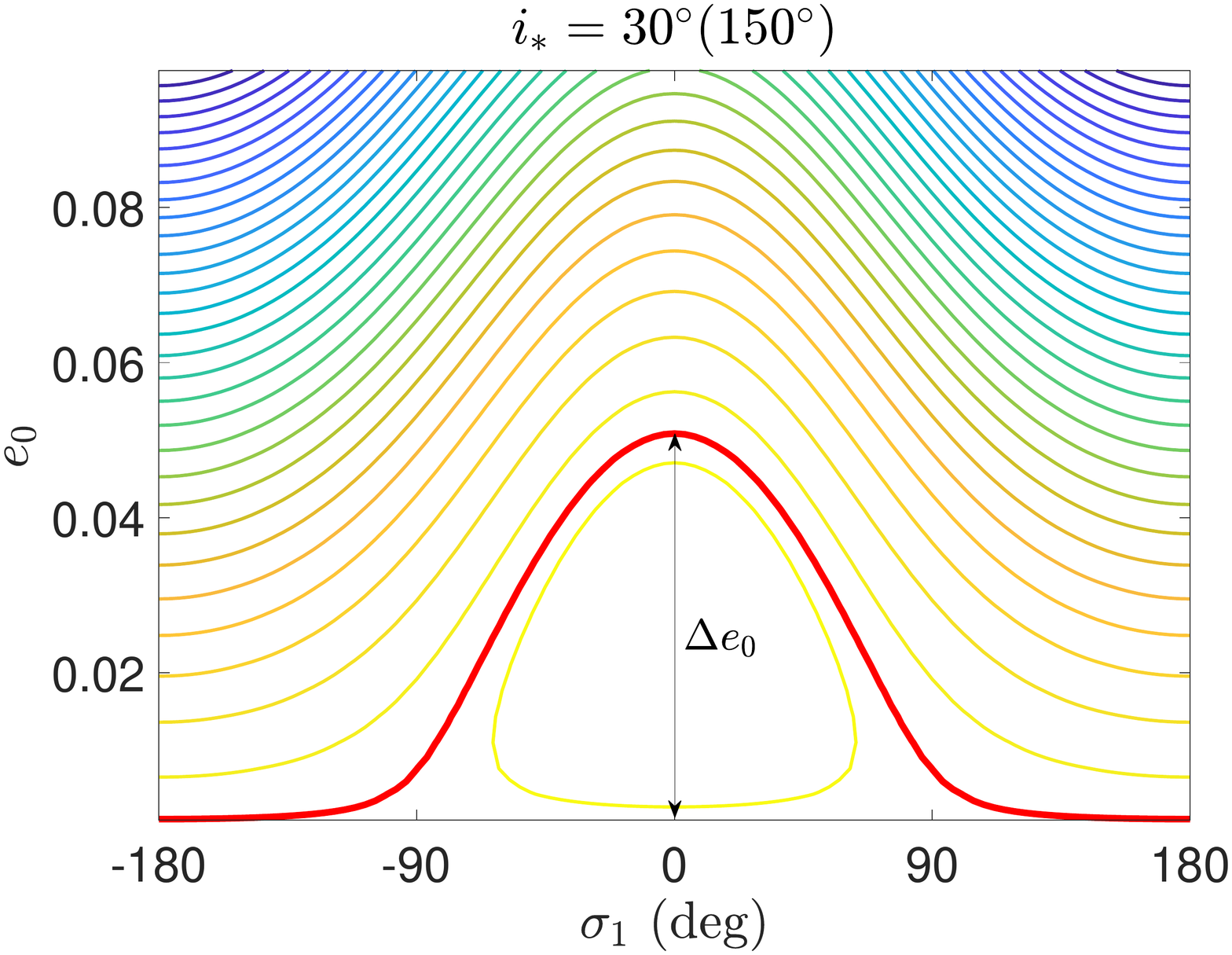}
\caption{Phase portraits (level curves of resonant Hamiltonian) for low-inclination cases, shown in the $(\sigma_1, e_0)$ space. Note that $e_0$ is the eccentricity when the angle $g$ is equal to zero. Dynamical separatrices are marked in red lines. The resonant width, denoted by $\Delta e_0$, measures the maximum size of island of libration.}
\label{Fig5}
\end{figure*}

When the critical inclination is at $i_* = 30^{\circ} (150^{\circ})$ (see the top-left panel of Fig. \ref{Fig4}), the structure arising in the phase portrait is pendulum-like: there is one resonant centre and one saddle point. The resonant centre is located at $(\sigma_1 = 180^{\circ}, i_0 = 90^{\circ})$ and the saddle point is located at $(\sigma_1 = 0^{\circ}, i_0 = 90^{\circ})$. The single island of libration is bounded by the dynamical separatrix shown by red line. As for this low-inclination case, there is another island of libration arising in low-eccentricity space, which will be shown in Fig. \ref{Fig5}.

When the critical inclination is increased up to $i_* = 45^{\circ} (135^{\circ})$ (see the top-right panel of Fig. \ref{Fig4}), the dynamical structures become complex. In total, there are three islands of libration: one is centred at $(\sigma_1 = 0, i_0 = 90^{\circ})$ and the other two islands are centred at $(\sigma_1 = 180^{\circ}, i_0 \ne 90^{\circ})$; the latter two islands of libration are symmetric with respect to $i_0 = 90^{\circ}$. There are three separatrices (shown by red lines), stemming from three saddle points. These separatrices provide boundaries for three isolated islands of libration in the phase space.

When the critical inclination is up to $i_* = 50^{\circ} (130^{\circ})$ (see the bottom-left panel of Fig. \ref{Fig4}), there are also three resonant centres and three saddle points. However, it is different from the case of $i_* = 45^{\circ} (135^{\circ})$ because there are only two separatrices: an inner separatrix and an outer separatrix. The inner separatrix bounds two asymmetric islands of libration, and the outer separatrix bounds the island centred at $(\sigma_1 = 0, i_0 = 90^{\circ})$ and the island centred at $(\sigma_1 = 180^{\circ},i_0 = 90^{\circ})$. The region bounded by the inner and outer separatrices is also of libration. The resonant trajectory inside this region holds the maximum variation of inclination up to $\sim$$80^{\circ}$.

When the critical inclination is at $i_* = 65^{\circ} (115^{\circ})$ (see the bottom-right panel of Fig. \ref{Fig4}), the dynamical structure becomes pendulum-like again: there is a single island of libration, centred at $(\sigma_1 = 180^{\circ}, i_0 = 90^{\circ})$. Along the resonant trajectory inside this island, the maximum variation of inclination can reach $\sim$$50^{\circ}$.

The phase portraits shown by Fig. \ref{Fig4} indicate that the inclinations of test particles can be effectively excited following along the trajectories inside islands of libration. Such a significant variation of orbit orientation is due to the effect of apsidal resonance under the octupole-order Hamiltonian model.

For the low-inclination cases, two examples are presented in Fig. \ref{Fig5} for the cases of $i_* = 20^{\circ} (160^{\circ})$ and $i_* = 30^{\circ} (150^{\circ})$. Different from Fig. \ref{Fig4}, the phase portraits are plotted in the $(\sigma_1, e_0)$ space. Here $e_0$ corresponds to the eccentricity when the angle $g$ is equal to zero. Besides the islands shown in Fig. \ref{Fig4} (see the top-left panel), a new island arises in the low-eccentricity space and is centred at $\sigma_1 = 0$. Also, the dynamical separatrices are shown in red lines. In these two plots, the resonant width is measured by $\Delta e_0$. Following along the trajectories inside islands of libration, the eccentricities of test particles can be excited due to the effect of apsidal resonance.

\begin{figure*}
\centering
\includegraphics[width=0.8\textwidth]{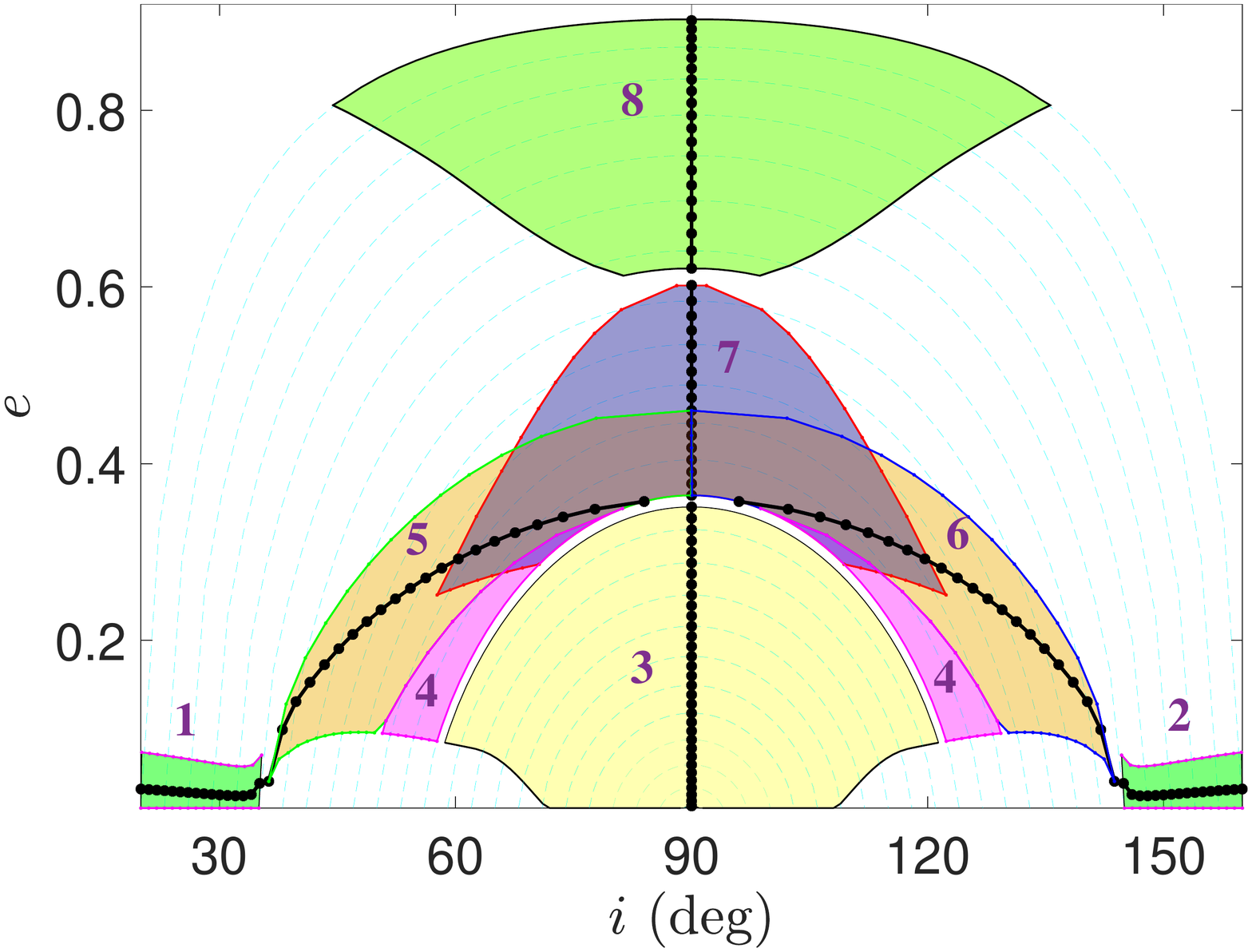}
\caption{Resonant centres of apsidal resonance distributed in the inclination--eccentricity space (black dots) and libration zones obtained by analysing phase portraits (shaded areas). The level curves of the motion integral $\Sigma_2$ are shown as background and the resonant width is measured along the isoline of $\Sigma_2$. Totally, there are eight libration zones, denoted by numbers from 1 to 8. Evidently, the distribution of libration zones is symmetric with respect to the line of $i=90^{\circ}$.}
\label{Fig6}
\end{figure*}

\subsection{Libration zones of apsidal resonance}
\label{Sect5_2}

By analysing phase portraits at different levels of motion integral $\Sigma_2$, it is possible for us to identify the location of resonant centre and boundaries of libration zones.

The main results of this work are reported in Fig. \ref{Fig6}, where the resonant centres are marked in black dots and libration zones are shown in shaded areas with different colors. Boundaries of each libration zone are provided by the dynamical separatrix evaluated at the angle of the corresponding resonant centre (see Figs. \ref{Fig4} and \ref{Fig5} for representative phase portraits). The level curves of the motion integral $\Sigma_2$ are plotted in dashed lines as background. It is noted that resonant width is measured along isolines of motion integral.

From Fig. \ref{Fig6}, we can see that dynamical structures arising in the $(e,i)$ space are symmetric with respect to the line of $i=90^{\circ}$. In total, there are five branches of libration centres: one branch located on the polar line, two branches occupying in the low-eccentricity and low-inclination space and the remaining two located in the space with eccentricities changing from $\sim$$0$ to $\sim$$0.4$. In addition, there are eight libration zones, denoted by numbers from 1 to 8. It should be noted that the branches of resonant centres in zones 1 and 2 are not present under the quadrupole-level dynamical model\footnote{See Fig. \ref{Fig3} for the nominal location of apsidal resonance.}. It means that these two branches are present due to the pure effect of octupole-order Hamiltonian. Except for the zones 1 and 2, all the other branches of libration centres are consistent with the ones shown in Fig. \ref{Fig3}.

As for libration zone 1 (see Fig. \ref{Fig5} for the representative phase portraits), the bottom boundary is located at $e=0$\footnote{It shows that zero-eccentricity points $e=0$ are saddle points of the associated resonant model.}, the resonant width decreases first and then increases with the inclination $i$. Inside this zone, \citet{funk2011influence} found an interesting dynamical region around $i$$\sim$$35^{\circ}$, where the low-eccentricity orbits are long-term stable, meaning that the inclined ($\sim$$35^{\circ}$) quasi-circular Earth-mass companion can be survived in the habitable zone of extrasolar system. \citet{libert2012interesting} explained that the long-term stable dynamics at inclination of $\sim$$35^{\circ}$ is due to the existence of the secular resonance associated with $\sigma = \omega - \Omega$. In the low-eccentricity space with $i<39^{\circ}$, \citet{lei2021} studied the long-term dynamics by means of Lie series transformation and they pointed out the long-term stability inside this zone is governed by the apsidal resonance with argument of $\sigma = \omega + \Omega$, which is consistent with the result of the current work.

Libration zone 1 has a symmetric zone in the retrograde space, denoted by 2. As for this zone, the bottom boundary of libration is located at $e=0$ and the resonant width decreases first and then increases with $\pi-i$. Inside zones 1 and 2, the effect of apsidal resonance is to excite the eccentricities. However, the inclination has little variation during the long-term evolution. This is the reason that the low-eccentricity islands of libration cannot be found in the phase portrait plotted in the $(\sigma_1, i_0)$ space (please refer to the first panel of Fig. \ref{Fig4}).

Libration zone 3 appears when the critical inclination is larger than $\sim$$59^{\circ}$ and smaller than $\sim$$121^{\circ}$ (see the bottom-right panel of Fig. \ref{Fig4} for the representative phase portrait). The eccentricity of this zone ranges from zero to $\sim$$0.35$. The effect of apsidal resonance inside this zone is to exchange the test particle's eccentricity and inclination. The resonant trajectories inside this zone hold resonant centres at $i=90^{\circ}$. Thus all the trajectories inside this zone could flip from prograde to retrograde and back again.

There are two subregions in zone 4: one in the prograde space and the other one in the retrograde space. According to the phase portrait shown in the bottom-left panel of Fig. \ref{Fig4}, we can see that this zone of libration holds centres at $i=90^{\circ}$ and it is bounded by the inner and outer separatrices. The effect of apsidal resonance inside this zone is to significantly change test particle's eccentricity and inclination. In addition, all the trajectories inside this zone could realise flips between prograde and retrograde.

Zones 5 and 6 are symmetric with respect to $i=90^{\circ}$. Please refer to the top-right panel of Fig. \ref{Fig4} for the representative phase portrait. We can see that, for these two zones, the line of $i=90^{\circ}$ provides one boundary, so that the resonant trajectories are restrained in either the prograde space or the retrograde space. It means that the trajectories inside these two zones cannot flip from prograde to retrograde or vice versa.

Let us move to the last two zones 7 and 8. Please refer to the top-left and -right panels of Fig. \ref{Fig4} for representative phase portraits. Zone 7 is located in the intermediate-eccentricity region, and zone 8 is located in the high-eccentricity region. Both of them hold resonant centres at $i=90^{\circ}$. As a result, all the resonant trajectories inside these two zones can flip from prograde to retrograde and back again.

\begin{figure*}
\centering
\includegraphics[width=0.4\textwidth]{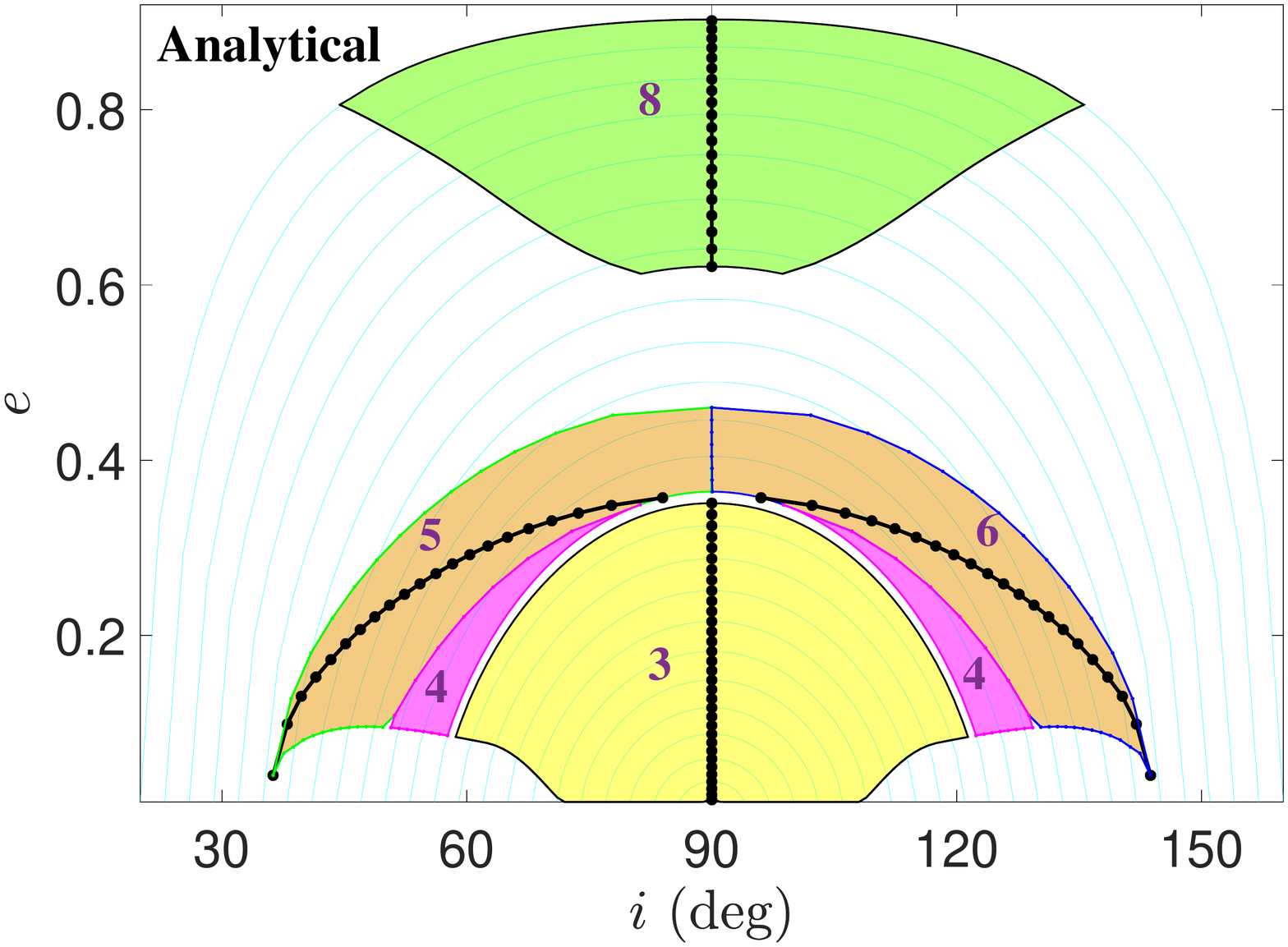}
\includegraphics[width=0.4\textwidth]{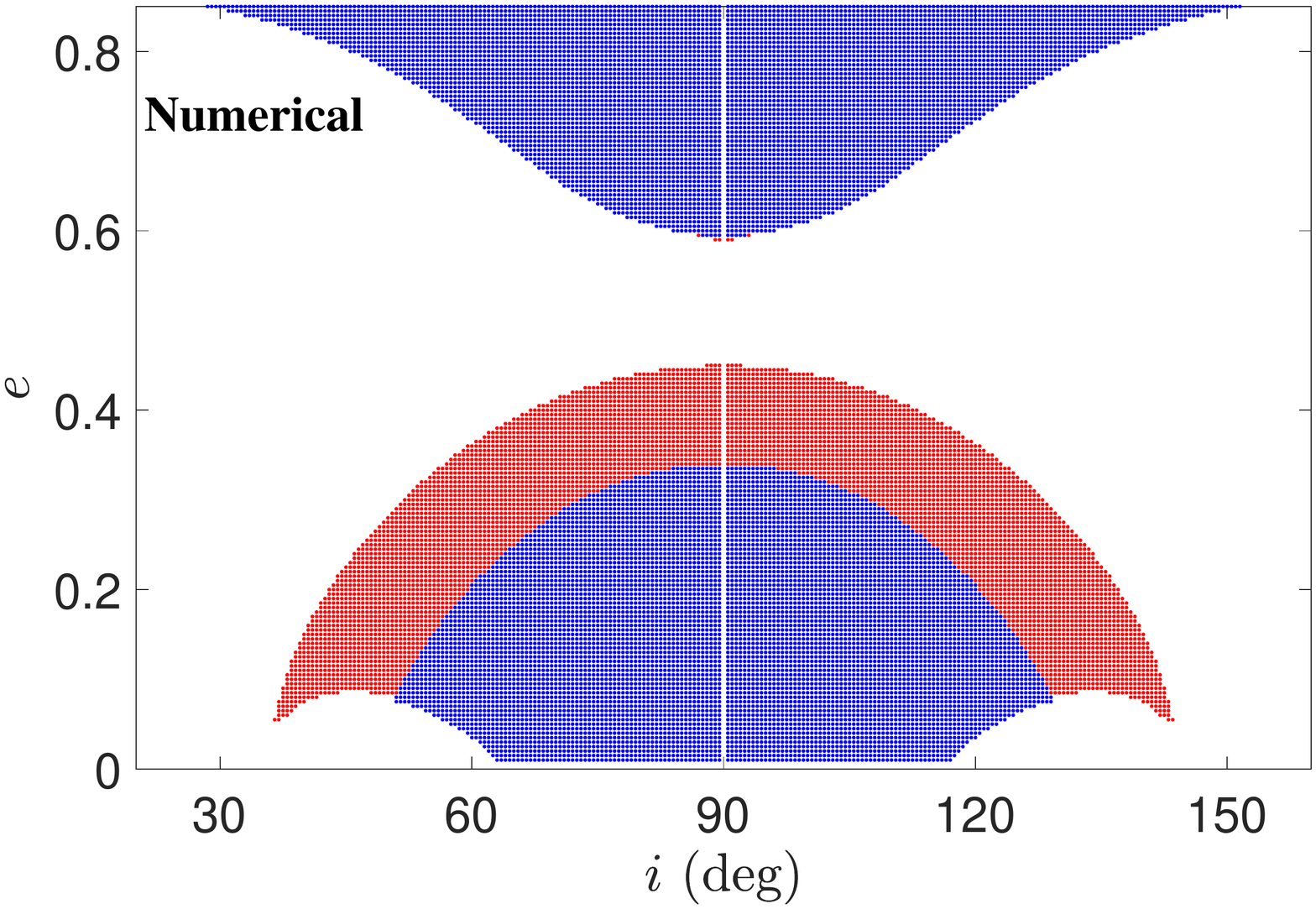}
\caption{Analytical results about the resonant regions for those apsidal resonances centred at $\sigma_{1,c} = \pi$ (\emph{left panel}) and the associated numerical results for the distribution of apsidal resonance (\emph{right panel}). For the numerical results, the initial condition is $g_0 = 0$ and $h_0 = \pi$ (corresponding to $\sigma_{1,c} = \pi$ at the initial instant). In the left panel, level curves of the motion integral $\Sigma_2$ are presented as background. In the right panel, blue dots stand for resonant trajectories inside islands centred at $i=90^{\circ}$ and red dots for resonant trajectories inside asymmetric islands of libration. It should be mentioned that the difference arising in the top space is due to the fact that the analytical results are restrained by level curves of the motion integral $\Sigma_2$.}
\label{Fig7}
\end{figure*}

\begin{figure*}
\centering
\includegraphics[width=0.4\textwidth]{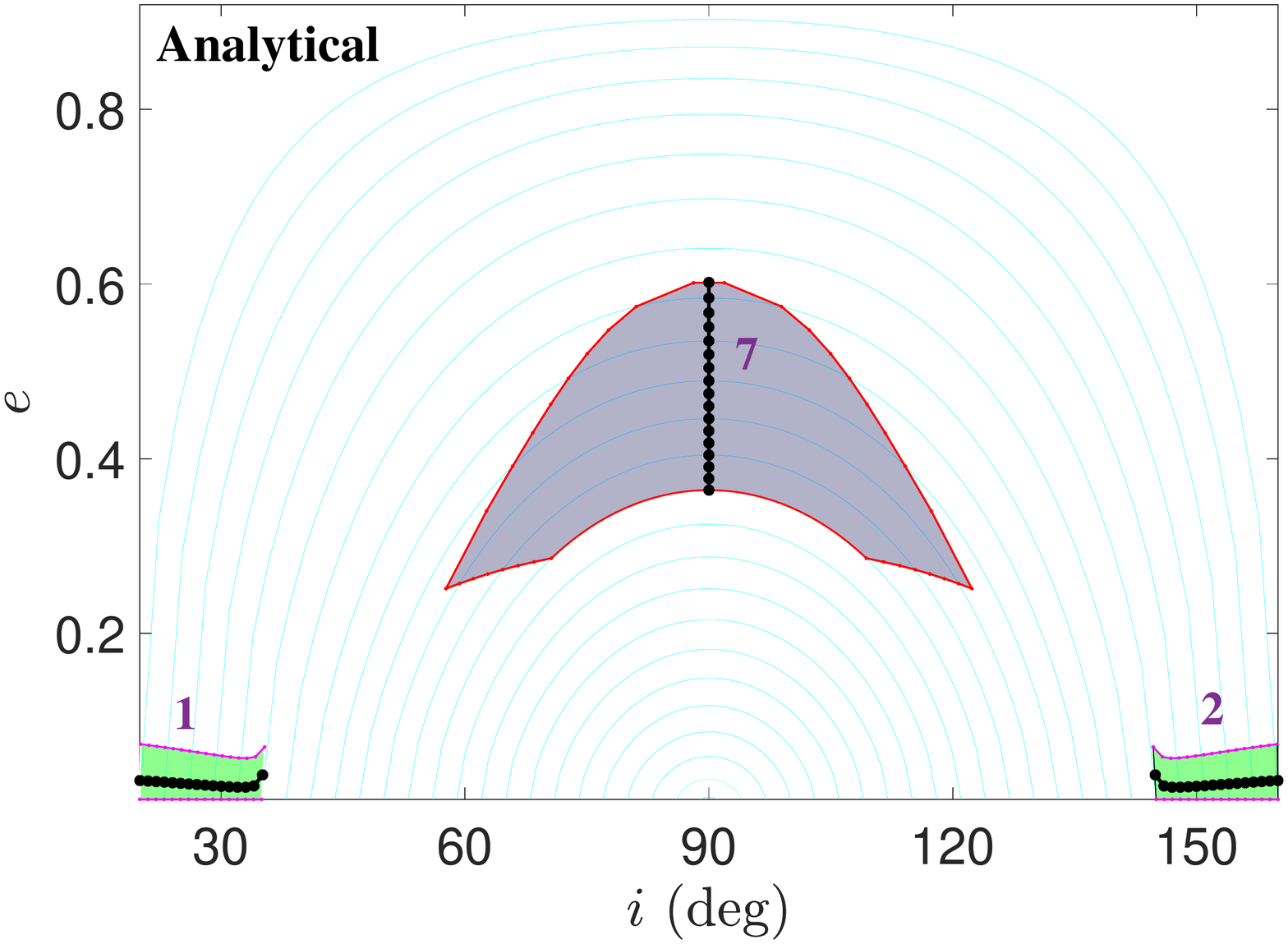}
\includegraphics[width=0.4\textwidth]{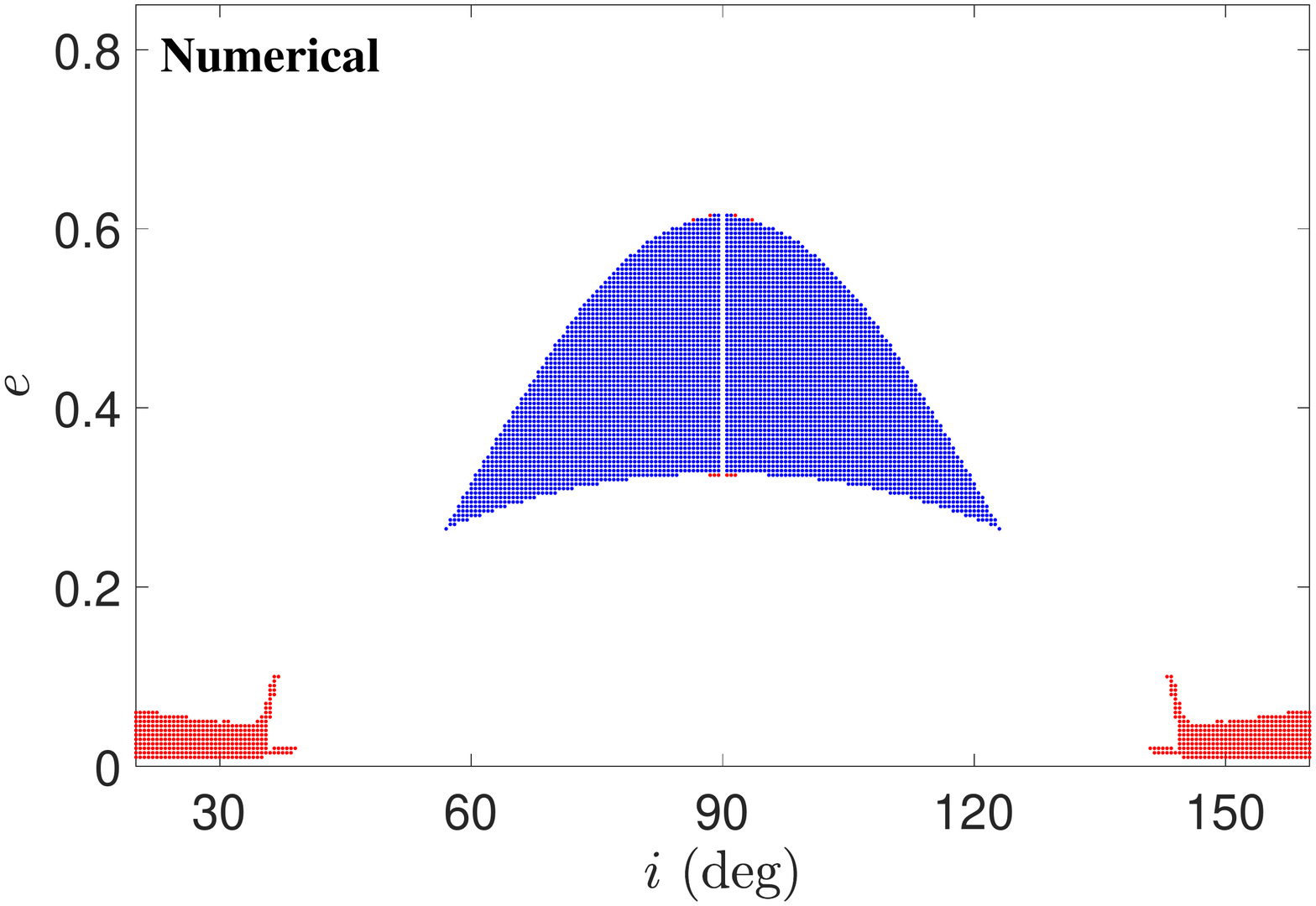}
\caption{Analytical results about the resonant regions for those apsidal resonances centred at $\sigma_{1,c} = 0$ (\emph{left panel}) and the associated numerical results for the distribution of apsidal resonance (\emph{right panel}). For the numerical results, the initial condition is $g_0 = 0$ and $h_0 = 0$ (corresponding to $\sigma_{1,c} = 0$ at the initial instant). In the left panel, level curves of the motion integral $\Sigma_2$ are presented as background. In the right panel, blue dots stand for resonant trajectories inside islands centred at $i=90^{\circ}$ and red dots for resonant trajectories occurring in low-eccentricity, low-inclination spaces.}
\label{Fig8}
\end{figure*}

Figures \ref{Fig7} and \ref{Fig8} provide comparisons between analytical and numerical results for the libration zones of apsidal resonance. In particular, Fig. \ref{Fig7} corresponds to apsidal resonances centred at $\sigma_{1,c} = \pi$ and Fig. \ref{Fig8} corresponds to the ones centred at $\sigma_{1,c} = 0$. For analytical results, libration zones 3, 4, 5, 6 and 8 are shown in the left panel of Fig. \ref{Fig7} and libration zones 1, 2 and 7 are presented in the left panel of Fig. \ref{Fig8}. To be consistent, the initial conditions of numerical results are assumed as $g_0 = \pi$ and $h_0 = 0$ corresponding to the resonant centre at $\sigma_{1,c} = \pi$, and they are assumed as $g_0 = 0$ and $h_0 = 0$ corresponding to the resonant centre at $\sigma_{1,c} = 0$. To produce numerical results, the equations of motion represented by Eq. (\ref{Eq3}) are numerically integrated over 500 units of dimensionless time. The orbit is recorded as a librating trajectory if the maximum variation of critical argument $\sigma = h + {\rm sign}(H)g$ is smaller than $2\pi$ during the considered integration period \footnote{This means that the critical argument is librating during the integration period.}. The numerical distributions of apsidal resonance are shown in the right panels of Figs. \ref{Fig7} and \ref{Fig8}. As expected, good agreement can be found between analytical and numerical results, indicating that the resonant Hamiltonian formulated in the previous section is valid and applicable to explore dynamics of apsidal resonance under the octupole-level approximation.

\subsection{Application to orbit flips}
\label{Sect5_3}

\begin{figure*}
\centering
\includegraphics[width=0.4\textwidth]{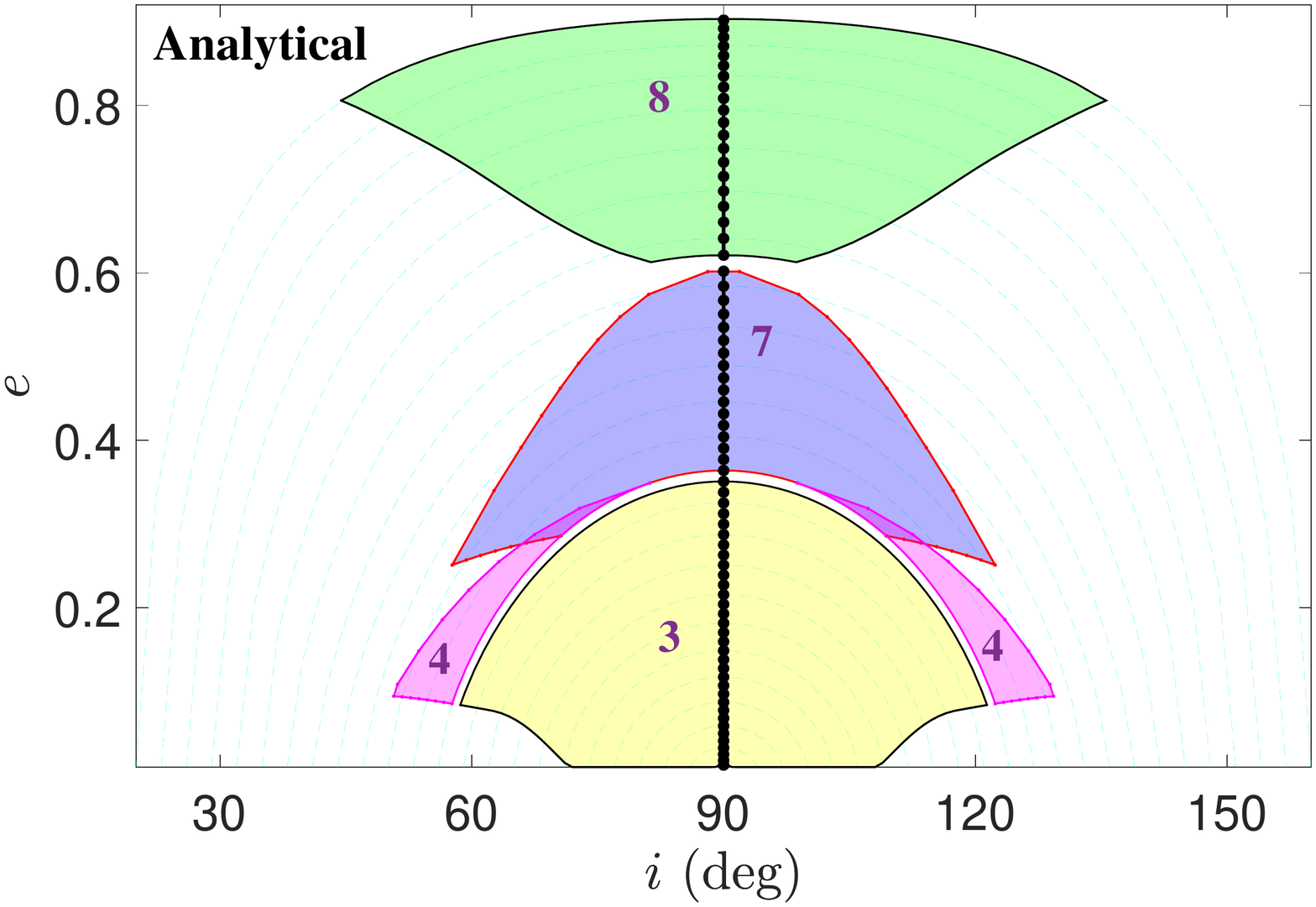}
\includegraphics[width=0.4\textwidth]{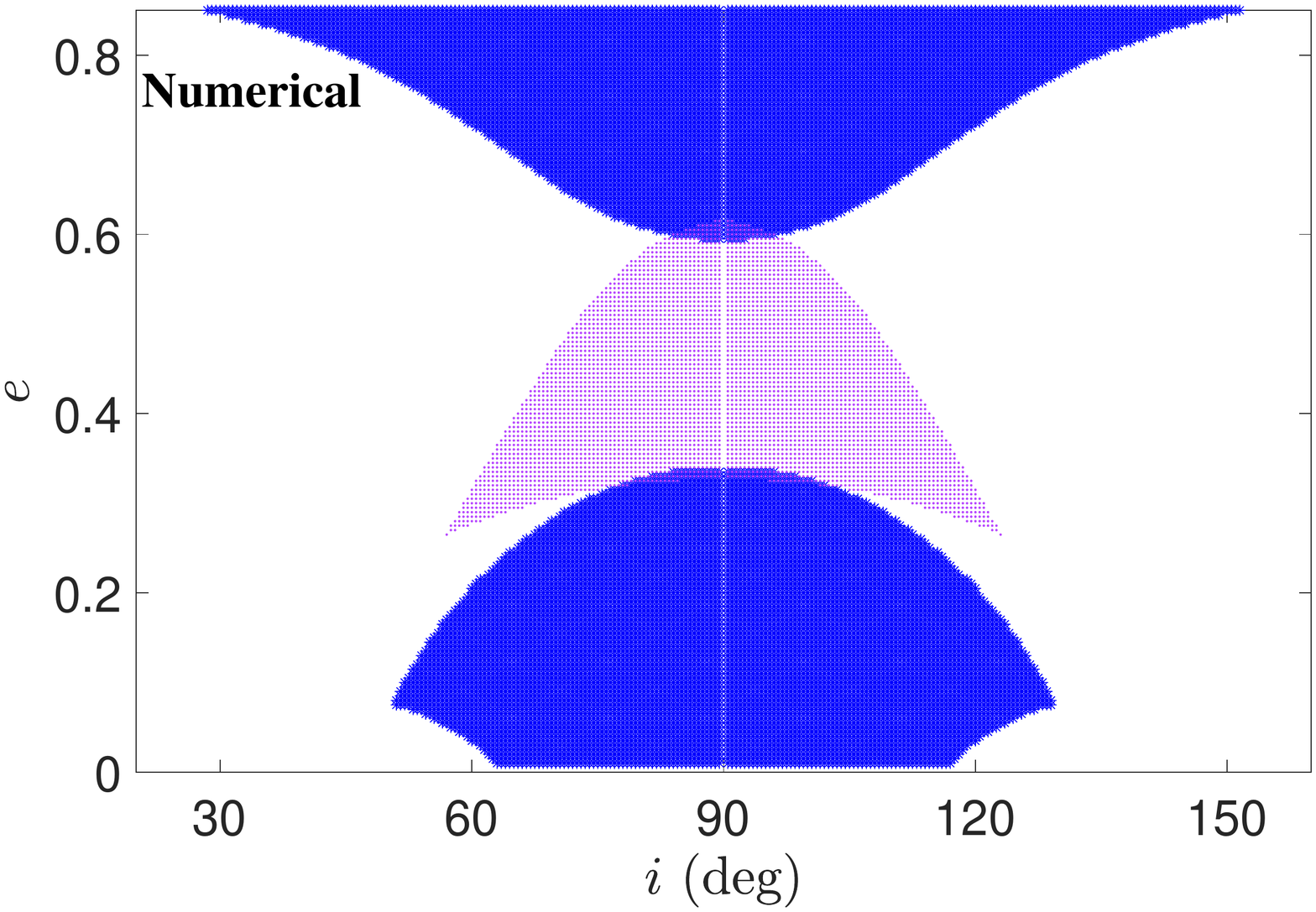}
\caption{Analytical results of libration zones for those apsidal resonances causing orbit flips (\emph{left panel}) and the numerical results of flipping regions (\emph{right panel}). In the left panel, there are four libration zones causing orbit flips (zones 3, 4, 7 and 8). It is observed that those apsidal resonances with centres at $i=90^{\circ}$ cause orbit flips. In the right panel, there are three distinct regions of orbit flips, located in the low-eccentricity, intermediate-eccentricity and high-eccentricity spaces.}
\label{Fig9}
\end{figure*}

According to the results given in Fig. \ref{Fig6}, it is known that the resonant trajectories with resonant centres at $i=90^{\circ}$ can flip from prograde to retrograde and back again. In Fig. \ref{Fig9}, analytical results of libration zones causing orbit flips are compared with the numerical distribution of flipping orbits under the octupole-level Hamiltonian model. In total, there are four libration zones causing orbit flips (zones 3, 4, 7 and 8). Inside zones 3, 4 and 8, the resonant centres are at $\sigma_{1,c} = \pi$ and, inside zone 7, the resonant centres are at $\sigma_{1,c}= 0$. Analytical results for the libration zones causing orbit flips are presented in the left panel of Fig. \ref{Fig9}.

To be consistent with analytical results, the initial conditions of numerical results are assumed at $g_0 = 0$ and $h_0 = 0$ for the case of $\sigma_{1,c}=0$ and they are assumed at $g_0 = 0$ and $h_0 = \pi$ for the case of $\sigma_{1,c}=\pi$. Similarly, the equations of motion are numerically integrated over 500 units of dimensionless time. The numerically propagated trajectories are recorded as flipping orbits if orbit inclinations can switch between prograde and retrograde. The numerical distribution of flipping orbits\footnote{Similar numerical results of flipping regions can be found in \citet{lei2022systematic} under the dynamical model specified by $\epsilon = 0.1$. In particular, the flipping region located in the low-eccentricity space corresponds to the LeHi case and the one located in the high-eccentricity space corresponds to the HeLi case shown in \citet{li2014eccentricity}.} is shown in the right panel of Fig. \ref{Fig9}.

From the right panel of Fig. \ref{Fig9}, it is observed that there are three distinct flipping regions in the $(i,e)$ space: one located in the low-eccentricity region, one located in the intermediate-eccentricity space and the third one located in the high-eccentricity space. The low-eccentricity region of orbit flips corresponds to libration zones 3 and 4. Inside such a low-eccentricity flipping region, the critical argument $\sigma = h + {\rm sign}(H) g$ is librating around $\pi$. The intermediate-eccentricity region of flipping orbits corresponds to libration zone 7, where the critical argument $\sigma = h + {\rm sign}(H) g$ is librating around $0$. At last, the high-eccentricity region of flipping orbits corresponds to libration zone 8, where the critical argument $\sigma = h + {\rm sign}(H) g$ is librating around $\pi$.

The comparison made in Fig. \ref{Fig9} shows an excellent agreement between analytical and numerical results. The results imply that the dynamics of orbit flips can be well understood with the help of dynamical structures of apsidal resonance.

\section{Conclusions}
\label{Sect6}

In this paper, the dynamics of apsidal resonance are studied by means of perturbation treatments under the octupole-level approximation in restricted hierarchial planetary systems.

The Hamiltonian function is composed of the quadrupole-order term ${\cal H}_2$ and the octupole-order term ${\cal H}_3$. From the viewpoint of perturbative treatments, the quadrupole-order Hamiltonian is considered as the kernel function, and the octupole-order Hamiltonian plays the role of perturbation to the quadrupole-order dynamics. By introducing the action-angle variables (a kind of canonical transformation), the quadrupole-order Hamiltonian can be converted to be independent on angular coordinates, leading to the fact that the action variables become conserved quantities under the quadrupole-order Hamiltonian flow. The transformed quadrupole-order Hamiltonian gives rise to the fundamental frequencies (or proper frequencies), which can be used to identify the nominal location of secular resonances. It is found that the secular resonances with critical argument of $\sigma = h^* +{\rm sign}(H^*)g^*$ happen in the considered parameter space. We have demonstrated that, in the test-particle limit, the argument $\sigma = h^* +{\rm sign}(H^*)g^*$ can be equivalently expressed as $\sigma = \varpi^* - \varpi_p$, which corresponds to apsidal resonances.

To study the dynamics of apsidal resonance, a canonical transformation is introduced. After transformation, it becomes a typical separable Hamiltonian model, so that we can apply first-order perturbation theory to formulate the resonant Hamiltonian model by averaging the Hamiltonian over rotating ZLK cycles. Application of first-order perturbation theory gives rise to a new constant of motion and the resulting resonant Hamiltonian model is of one degree of freedom. Phase portraits (level curves of resonant Hamiltonian with given motion integral) can be used to analyse the global dynamical structures of apsidal resonance. In particular, the location of resonance centre, saddle points, dynamical separatrix between circulating and librating regions as well as islands of libration can be determined from the resonant model.

Our main results are reported in Fig. \ref{Fig6}. It is concluded that (a) dynamical structures are symmetric with respect to $i=90^{\circ}$, (b) there are five branches of libration centres, and (c) there are eight libration zones. The comparisons between analytical and numerical results for libration zones of apsidal resonance shows good agreements between them.

It is found that islands of libration centred at $i=90^{\circ}$ can cause orbit flips. Thus, those libration zones with resonant centres at $i=90^{\circ}$ correspond to flipping regions in the phase space. To validate this point, the analytical results of libration zones are compared with numerical distributions of flipping orbits. A perfect correspondence can be found between the analytical and numerical results.

Through this study, we can conclude that, from the viewpoint of dynamics, the eccentric ZLK effect is equivalent to the effect of apsidal resonance at the octupole-level approximation in restricted hierarchical planetary systems. The dynamical response of the eccentric ZLK effect (or the effect of apsidal resonance) is to significantly excite eccentricities and/or inclinations (even flipping) of test particles in the very long-term evolution.

\section*{Acknowledgements}
This work is supported by the National Natural Science Foundation of China (Nos 12073011, 12073019).

%\begin{appendix}
%
%\section{Applications to other TNOs}
%\label{Sect_A1}
%
%\end{appendix}


\bibliographystyle{aa}
\bibliography{mybib}

%\begin{thebibliography}{}

%\end{thebibliography}

\end{document}